\documentclass[journal,comsoc]{IEEEtran}
\IEEEoverridecommandlockouts
\usepackage{cite}
\usepackage{amsmath, amssymb, amsfonts}
\usepackage{tabularx}
\usepackage{hyperref}
\usepackage{gensymb}
\usepackage{algorithmic}
\usepackage{array}
\usepackage[caption=false,font=normalsize,labelfont=sf,textfont=sf]{subfig}
\usepackage{textcomp}
\usepackage{stfloats}
\usepackage{url}
\usepackage{verbatim}
\usepackage{graphicx}
\def\BibTeX{{\rm B\kern-.05em{\sc i\kern-.025em b}\kern-.08em
    T\kern-.1667em\lower.7ex\hbox{E}\kern-.125emX}}
\usepackage{balance}

\begin{document}

\title{A Human-Centric Metaverse Enabled by Brain-Computer Interface: A Survey
\author{\IEEEauthorblockN{Howe Yuan Zhu, Nguyen Quang Hieu, Dinh Thai Hoang, Diep N. Nguyen, and Chin-Teng Lin}}

\thanks{Howe Zhu, Nguyen Quang Hieu, Dinh Thai Hoang, Diep N. Nguyen, and CT Lin are with 
University of Technology Sydney, Ultimo, NSW 2007, AU (e-mail: howe.zhu@uts.edu.au, hieu.nguyen-1@student.uts.edu.au, hoang.dinh@uts.edu.au, diep.nguyen@uts.edu.au, and chin-teng.lin@uts.edu.au).}
}

\markboth{IEEE Communications Surveys \& Tutorials,~Vol.~00, No.~00, September~2023}%
{IEEE Communications Surveys \& Tutorials,~Vol.~00, No.~00, September~2023}

\maketitle

\begin{abstract}
    The growing interest in the Metaverse has generated momentum for members of academia and industry to innovate toward realizing the Metaverse world. The Metaverse is a unique, continuous, and shared virtual world where humans embody a digital form within an online platform. Through a digital avatar, Metaverse users should have a perceptual presence within the environment and can interact and control the virtual world around them. Thus, a human-centric design is a crucial element of the Metaverse. The human users are not only the central entity but also the source of multi-sensory data that can be used to enrich the Metaverse ecosystem. In this survey, we study the potential applications of Brain-Computer Interface (BCI) technologies that can enhance the experience of Metaverse users. By directly communicating with the human brain, the most complex organ in the human body, BCI technologies hold the potential for the most intuitive human-machine system operating at the speed of thought. BCI technologies can enable various innovative applications for the Metaverse through this neural pathway, such as user cognitive state monitoring, digital avatar control, virtual interactions, and imagined speech communications. This survey first outlines the fundamental background of the Metaverse and BCI technologies. We then discuss the current challenges of the Metaverse that can potentially be addressed by BCI, such as motion sickness when users experience virtual environments or the negative emotional states of users in immersive virtual applications. After that, we propose and discuss a new research direction called Human Digital Twin, in which digital twins can create an intelligent and interactable avatar from the user's brain signals. We also present the challenges and potential solutions in synchronizing and communicating between virtual and physical entities in the Metaverse. Finally, we highlight the challenges, open issues, and future research directions for BCI-enabled Metaverse systems.
\end{abstract}


\section{Introduction and Motivations}
\label{sec:introduction}
\subsection{The Development of a Human-Centric Metaverse}
\begin{figure*}[t]
\centering
\includegraphics[width=0.7\linewidth]{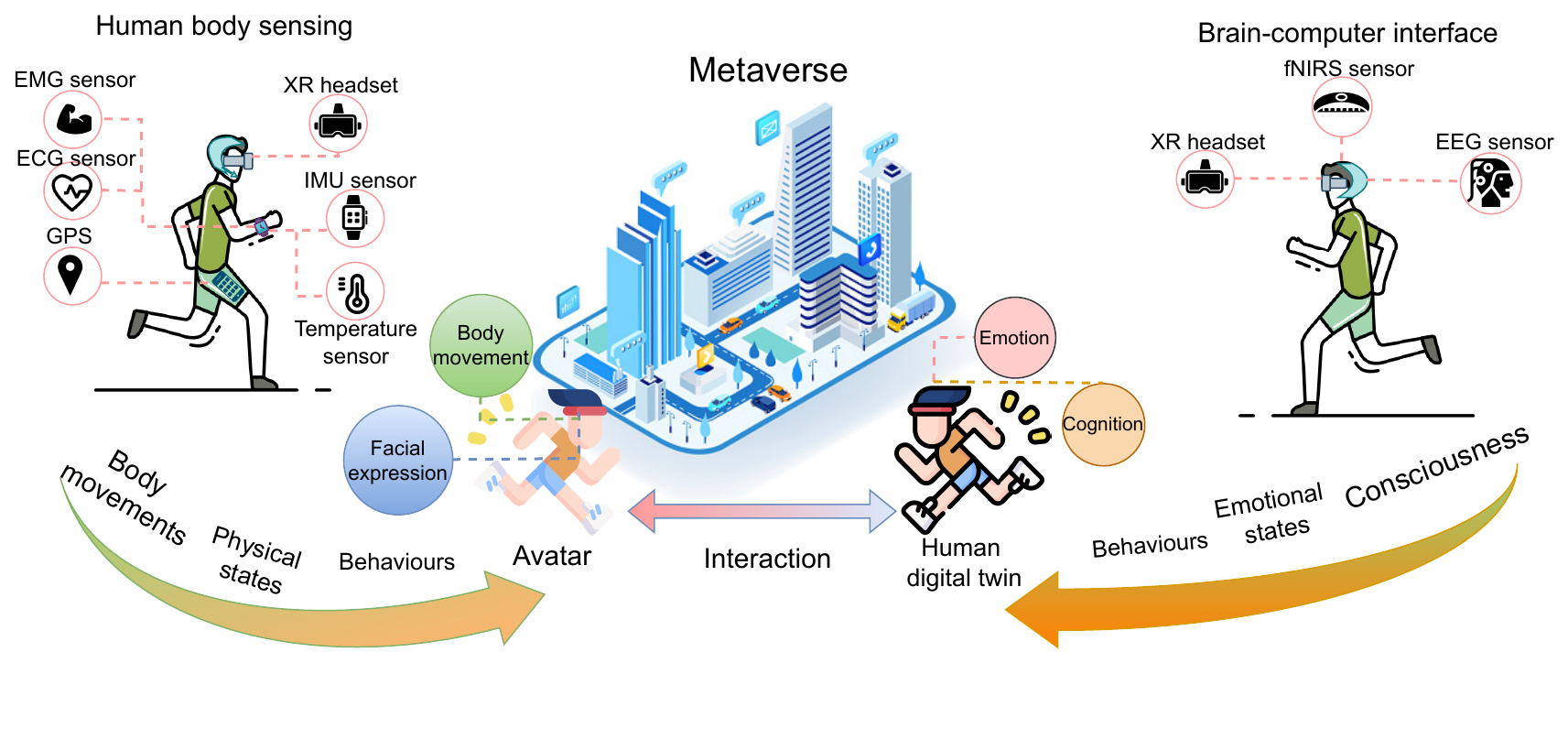}
\caption{An illustration of a human-centric design for the Metaverse. Users' information can be collected using human body sensing techniques, e.g., heart rate sensors, VR headsets, inertial measurement units (IMUs), and motion capture systems. Human characteristics such as body shapes, facial expressions, behaviours, and health can be revealed by analyzing the collected data. Unlike conventional human body sensing approaches, BCI can provide an alternative method to create intelligent avatars from a singular data source, i.e., human digital twins. Additional information from human users, e.g., emotional state and motion sickness, can be transferred to their avatars through BCI.
}
\label{fig:human-centric-metaverse}
\end{figure*}
\IEEEPARstart{T}{he} term ``Metaverse" was first coined by Neal Stephenson in his science fiction novel ``Snow Crash" in 1992. In this book, Stephenson described the Metaverse as a parallel existence of the physical and virtual worlds where users can interact with each other through their avatars. Although the idea of the Metaverse emerged 30 years ago, the technology was not yet ready for it until recent years. Recently, a large number of research and innovation efforts have been put into developing virtual reality (VR), extended reality (XR), and computing services to bring the Metaverse closer to reality \cite{xu2022full}. Previously, VR and XR technologies have developed the first building blocks for the Metaverse, such as 3D video games, VR games, and later mobile XR games like Pokemon Go. The development of the Metaverse beyond games and social media platforms has begun to realize it as the next generation of the Internet. Recent works and products have explored Metaverse applications, such as healthcare \cite{yu2012building}, e-commerce \cite{jeong2022innovative}, entertainment, and education \cite{ng2021optimal}. These ventures have shown great potential for revenue growth in the Metaverse, even though the Metaverse has not yet been fully realized.

As the Metaverse is still in its early stages, many ongoing parallel multifaceted research areas and challenges, including VR/XR, human-machine interfaces, and computing services, require substantial investment and development. Recent works and surveys have attempted to define architectures for the Metaverse based on these components. One particular area is exploring a human-centric design to enhance users' experience in the Metaverse \cite{lee2021all}. A human-centric design is a method of utilizing a user's behavioral, psychological, physiological, and observation information to improve the performance and usability of a system \cite{Morrissey1998}. In principle, a human-centric approach is to design an intuitive system by incorporating the user's potential state and modes of interaction. However, current human interaction methods, such as computer mouse clicks and keyboards, may not be intuitive within the new Metaverse experience.

Specifically, in Metaverse, the users can explore their surrounding environments using the control and sensory feedback from their hands, eyes, or thoughts in an immersive virtual environment \cite{wang2022survey, lee2021all}. Although a large body of work has been proposed recently, the human-machine interfaces are lagging compared to other aspects of the Metaverse. However, human-machine interaction is the primary channel that links the human body, the center of the Metaverse, to machines, i.e., any supporting devices and infrastructures. Conventional sensing techniques such as radio sensing, cameras, and wireless sensors can be utilized to develop a human-machine interface in the Metaverse. For example, face tracking, eye tracking, photogrammetry, computer vision, and motion capture can be used to construct fully immersive avatars in the Metaverse \cite{winkler2022questsim, yi2022physical, raj2021pixel}. In Fig.~\ref{fig:human-centric-metaverse}, we illustrate a human-centric design for the Metaverse in which we focus on the interaction aspect of the users between the physical world and the virtual world. For this, the virtual world can be enriched by using the cognitive interactions of the users. Such cognitive interactions can be collected through sensing techniques such as eye tracking, voice detection, and computer vision.

Using cognitive interactions of the users, or biological signals of the users in general, has shown its potential in designing and developing human-centric VR applications. For example, eye movement and heart rate data have been widely used in several applications, ranging from healthcare to robotics and virtual reality control \cite{tieri2018virtual}. Early findings showed that brain signals are encoded with a higher fidelity of sensory information than conventional sensing techniques, such as heart rate measurement and eye tracking. Toward this vision, Brain-Computer Interfaces (BCIs) have been considered in such applications as a neural interface between users and applications. As an attractive research area for exploiting human cognition in enabling technologies, BCI is inevitably becoming a part of the Metaverse.

\subsection{Towards a Human-Centric Metaverse using BCI}
This paper aims to provide a comprehensive survey about using BCI to enable a human-centric design for the Metaverse and discuss its associated communications challenges and future research directions/opportunities. BCI offers the ability to monitor the state of a Metaverse user and facilitate intuitive modes of user interaction within the Metaverse. BCI research began in 1875, Richard Canton, a British physicist, discovered the existence of electrical signals in the brains of animals. Only four decades later, Han Berger, a psychiatrist, invented the first measurement device allowing humans to measure the brain's electrical activity for the first time \cite{haas_2003}. After decades of research, BCI technologies exceeded their original scope in clinical trials and started attracting attention from the industry. Since the re-emergence of the Metaverse, researchers and industry have early adopted BCI as an enabling technology for the Metaverse \cite{xu2022full}. BCI can provide rich information from brain activity for building virtual environments and digital avatars beyond conventional sensing approaches such as eye movement tracking, touch, and audible sensors. We believe that BCI technology enables the following unique opportunities in Metaverse that would be unachievable with conventional sensing:

\begin{itemize}
\item \textbf{Direct communication to the brain: }BCI devices offer the unique ability to bypass the peripheral motor-sensory nervous/bodily systems to communicate directly with the brain. The brain is the complex motor-sensory control center of the body. The capability to interface with the brain enables the real-time reading of motor actions before execution and the response to sensory information, as it is processed within the brain \cite{ienca2016hacking}. This also enables the Metaverse users to send voluntary and directed commands for communication and control, adding additional channels that convey highly relevant information about the users' intent \cite{zander2011towards}. BCI can enable Metaverse users to relay complex commands, such as locomotion, limb movement, speech, and planned actions, to the Metaverse.

\item  \textbf{Multimodal information encoded onto a singular sensor: }The brain serves as the body's central nexus for motor-sensory information. A distinctive aspect of BCI sensors is the ability to capture the brain's complex neurological activity. The data acquired from these BCI sensors offer rich and multimodal motor-sensory information condensed into a singular sensor signal \cite{ienca2016hacking}. These signals can encompass various cognitive processes, sensory stimuli, emotions, and motor functions. Within a Metaverse BCI system, this information can be harnessed and translated across various brain activities, including perception, attention, memory, language processing, motor control, and emotional states.

\item \textbf{Higher degrees of encoded information within the signal: }The degree of encoded information is a crucial drawback of conventional peripheral biomonitor sensors, such as cardiovascular, electrodermal activity, and motion tracking \cite{zhu2021}. These sensors can reliably detect discrete changes in cognitive, mental and emotional states. However, they are less sensitive to transient or subtle shifts \cite{zhu2022effects}. In contrast, the neurological signals measured on BCI sensors possess a higher degree of encoded information that can be used to accurately measure subtle changes in cognitive, mental, and emotional states \cite{John2022}. Additionally, BCI signals can enhance our understanding of complex neurological states that cannot be measured through conventional sensing methods \cite{Do2021}. This capability can significantly enrich the understanding of the cognitive, mental, and emotional state of Metaverse users and enhance their personal experiences.

\item  \textbf{Intuitive control and natural interaction: }BCI can tap into the user's intentions, thoughts, and cognitive states. Instead of relying on overt physical actions, such as pressing buttons on a controller or moving a mouse, BCI can interpret the user's internal mental states. BCI can be a supportive mechanism to convey user intentions and desires without needing explicit external movements, resulting in a more intuitive and seamless control experience \cite{aldini2019effect,singh2020prediction}. This BCI application can benefit individuals with limited or impaired motor function \cite{lin2018wireless}. BCI can aid these individuals to regain control and interact with their environment using intact brain activity. In the Metaverse, an intuitive/natural user interface can enable a Metaverse avatar that acts as the virtual prosthesis/extension of a Metaverse user.  

\item  \textbf{Universal Access: }A valuable aspect of BCI technology is the ability to restore/replace motor functions for individuals with motor impairments, such as paralysis or limb loss \cite{lin2018wireless}. BCI technologies can bypass traditional motor pathways and allow individuals to directly interface with various assistive technologies to enable interaction with their environments using their brain activities. BCI has the potential to significantly enhance the quality of life and enable individuals to perform tasks that would otherwise be challenging or impossible. BCI can also be integrated with other assistive technologies or input modalities to create multimodal interfaces \cite{panoulas2010brain}. If actualized, BCI can enable universal access for Metaverse users with disabilities. Within the Metaverse, users can experience all the features of the virtual environment without restrictions on physical or mental capabilities.  
\end{itemize}

While there are several advantages to BCI technology, there are also several challenges hindering the integration of BCI for Metaverse users. The first challenge in the Metaverse is the construction of virtual embodiments, which involve multi-sensory data from the Metaverse users. Conventional user embodiment and user interaction schemes require many wearable and external sensing devices, such as handheld controllers for user input, wearable trackers for body kinematics, wearable cameras on head-mounted devices as facial sensors, heart rate trackers for workload and emotion measurement, and pupil cameras for eye movement tracking. Each sensing technique requires specific hardware devices, sensors, and customized software to serve a particular application, limiting the scalability and synchronization of multi-sensory data sources. In this case, BCI can act as a neural interface that integrates multiple sensing modalities, such as limb movement, intentions, emotion, and eye movement, into a single wearable signal source \cite{tidoni2014audio, cheng2020brain, wu2017evaluation}. The second challenge is a lack of individualization for the Metaverse technology. The multiple sensing modalities of conventional technologies also raise significant concerns about utilizing multi-sensory data sources to tailor applications to individual needs. As the sensing data come from different sources, it is challenging to fully utilize, synchronize, and distil information from different modalities, such as emotion recognition and eye movement \cite{zhang2021survey}.
Unlike traditional approaches, BCI offers an information source that can be individualized and tailored to the experience of individuals. The third challenge is the real-time or near real-time processing, and communications of BCI signals in Metaverse \cite{ning2021survey}. A critical factor is the scalability of the computation load and power for the large user base within the Metaverse. The current communications and computing capabilities will be insufficient to actualize the full-scale of the Metaverse. 
The fourth challenge is our limited knowledge of virtual embodiment, as we have not been able to fully actualize a virtual being/avatar in the Metaverse world. To address this challenge, we discuss the potential of a human digital twin (HDT) solution to provide a viable solution and open emerging applications for the Metaverse using BCI.

\subsection{Related Surveys and Our Contributions}

Various surveys have been conducted on the Metaverse, covering architecture, applications, technologies, security, and privacy concerns. In \cite{lee2021all}, the authors discussed the potential architecture and applications for the Metaverse. In \cite{xu2022full}, the authors examined the enabling technologies for the Metaverse from a cloud/edge computing perspective. In \cite{pham2022artificial}, the authors analyzed the potential applications of machine learning and deep learning for the Metaverse. In \cite{wang2022survey}, privacy and security issues of the Metaverse were discussed in detail. These existing surveys about the Metaverse focus on general issues and technologies, often overlooking the incorporation of human factors within the Metaverse. Although the authors in \cite{pham2022artificial} and \cite{wang2022survey} discussed the idea of using BCI as a neural interface between users and the Metaverse, none of them examined the BCI techniques involved in decoding human behavior. In \cite{bernal2022brain}, the authors considered potential BCI applications for the Metaverse. However, the context of the work was limited to the surface of BCI and the Metaverse. For example, the authors did not address the applications, limitations, and challenges, such as synchronizing and communicating between entities, let alone applying them in the Metaverse. Overall, BCI was not discussed in detail in the aforementioned surveys. To the best of our knowledge, our survey is the first in the literature to comprehensively discuss BCI's potential in the Metaverse.

In particular, in this survey, we aim to provide a comprehensive survey about BCI technologies and their potential for future development of the Metaverse. Furthermore, we propose and discuss a new concept of Human Digital Twin (HDT), a new approach to constructing human embodiment within the Metaverse using BCI. In summary, our main contributions are as follows:
\begin{itemize}
\item We provide the fundamental background and discuss the current challenges of the Metaverse that conventional sensing approaches could not effectively address.
\item We provide the background of BCI, focusing mainly on non-invasive BCI, which is more suitable for commercial applications. We then describe how BCI technologies can enhance user embodiment within the Metaverse. We also review BCI-enabled interaction schemes for the Metaverse users and describe the differences between the BCI sensing technologies and the conventional sensing techniques.
\item We introduce the new concept of HDT. With HDT, we can develop individualized Metaverse applications and enhance our knowledge of virtual embodiment in the Metaverse. We further discuss potential challenges for integrating HDT in the Metaverse, such as real-time communications, synchronization, and interactions between HDTs.
\item We highlight the challenges, open issues, and future research directions of BCI technologies for the Metaverse. The ethics and security of using BCI for the Metaverse are also discussed. Alternatively, the potential applications, such as brain communications, are also discussed.
\end{itemize}

\subsection{Paper Organization}
\begin{figure}[t]
    \centering
    \includegraphics[width=1.0\linewidth]{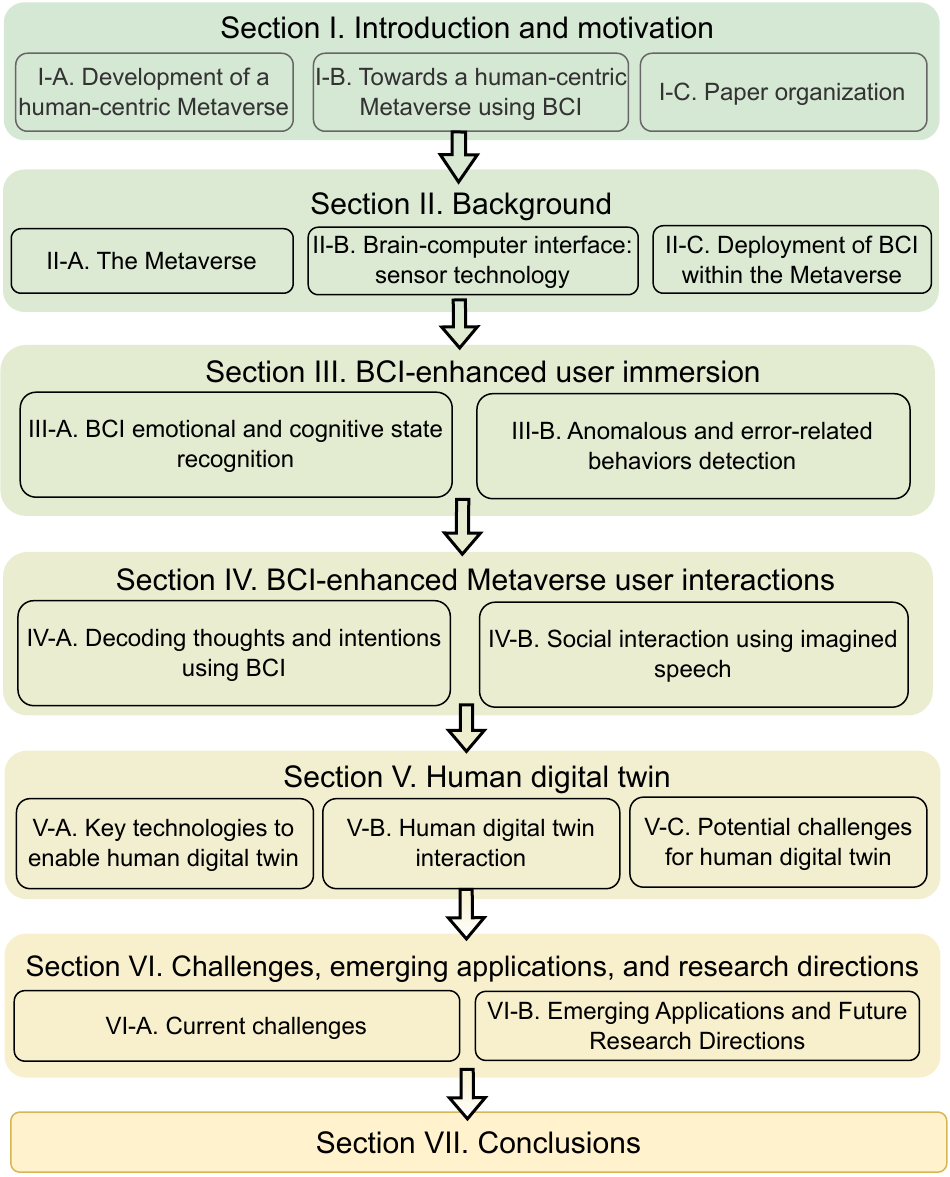}
    \caption{The organization of the survey.}   
    \label{fig:organization}
\end{figure}
As illustrated in Fig.~\ref{fig:organization}, our paper is organized as follows.
Section \ref{sec:background} provides the background of the Metaverse and BCI technologies.
Section \ref{sec:bci-enhanced-embodiment} provides more details about BCI technologies, including emotional and cognitive state recognition for the Metaverse.
This section further discusses the potential approaches to prevent error-related behaviors, such as VR motion sickness, stress, and fatigue, in the Metaverse.
Section \ref{sec:bci-enabled-metaverse} describes the potential interactions between the users and the Metaverse through BCI.
This section also discusses VR-BCI user interface design paradigms for the Metaverse applications.
In Section \ref{sec:human-digital-twin}, we propose a new concept of the HDT in which the digital twin is utilized to create twin entities for human users from their brain signals.
We also discuss challenges in synchronizing and real-time communication between virtual and physical entities.
In Section \ref{sec:open-issues}, we highlight the current challenges, open issues, and future research directions toward BCI-enabled Metaverse systems.
This section also discusses ethics, security, privacy issues, and emerging applications, including hardware, software, and algorithm designs.
Finally, Section \ref{sec:conclusion} concludes the paper. 
In addition, we provide a list of abbreviations and descriptions used in this paper in Table \ref{tab:Abb}.

\begin{table*}[t]
	\footnotesize
	\centering
	\caption{\footnotesize LIST OF COMMON ABBREVIATIONS USED IN THIS SURVEY} \label{tab:abbreviation} 
	\begin{tabular}{|>{\raggedright\arraybackslash}m{1.5cm}|>{\raggedright\arraybackslash}m{6cm}|>{\raggedright\arraybackslash}m{1.5cm}|>{\raggedright\arraybackslash}m{6cm}|}
		\hline
		\multicolumn{1}{|>{\centering\arraybackslash}m{1.5cm}|}{\textbf{Abbreviation}} & \multicolumn{1}{>{\centering\arraybackslash}m{6cm}|}{\textbf{Description}} & \multicolumn{1}{>{\centering\arraybackslash}m{1.5cm}|}{\textbf{Abbreviation}} & \multicolumn{1}{>{\centering\arraybackslash}m{6cm}|}{\textbf{Description}} \\ \hline		\hline
		HDT & Human Digital Twin & VR & Virtual Reality  \\ \hline
		XR & Extended Reality & EEG & Electroencephalogram \\ \hline
		DoF & Degree-of-Freedom  & AI & Artificial Intelligence \\ \hline
		IoT & Internet of Things & QoE & Quality-of-Experience \\ \hline
		ECoG & Electrocorticography & MEG & Magnetoencephalogram \\ \hline
		fNIRS & Functional Near-Infrared Spectroscopy & fMRI & Functional Magnetic Resonance Imaging\\ \hline
		FOV & Field of View & HMD &  Head-mounted Display\\ \hline
        ERN & Error-related Negativity & MI &  Motor Imagery\\ \hline
		SSVEP & Steady-State Visual Evoked Potential & TDMA & Time Division Multiple Access \\ \hline
		MISO & Multiple Input Single Output & MIMO  & Multiple Input Multiple Output \\ \hline
		SDMA & Spatial Division Multiple Access & NOMA & Non-Orthogonal Multiple Access \\ \hline
		RSMA & Rate-Splitting Multiple Access & MMO & Massive Multiplayer Online (a video game genre) \\ \hline
	\end{tabular} 
 \label{tab:Abb}
\end{table*}

\section{Background}
\label{sec:background}
\subsection{The Metaverse}

The term ``Metaverse" is a combination of the prefix ``Meta" (meaning beyond) and the suffix ``verse" (meaning universe). As its name suggests, the Metaverse is a universe of the next-generation Internet that allows the parallel existence of the physical world and shares 3D virtual worlds. The earliest concepts of the Metaverse can be found in classical MMO (Massively multiplayer online) games \cite{Nevelsteen2017}. In these games, users are given a uniquely singular persistent virtual world as a medium for social and worldly interactions. The Metaverse shifts this paradigm by incorporating modern technology to generate a seamlessly immersive experience transitioning between the physical and virtual worlds. The envisioned Metaverse is where users can naturally (touch the environment with their hand or walk with their feet) move and interact with the virtual environment as though they are in the physical world \cite{lee2021all}. Unlike conventional interactions in the current Internet, where we use devices such as a mouse, cursor, and keyboard, the Metaverse enables users to immerse applications and services through their digital avatars with supporting VR and XR technologies. As a result, users in the Metaverse can potentially eliminate Spatio-temporal barriers in how they work, live, and entertain. To this end, the Metaverse can be developed from the convergence of multiple supporting engines such as VR/XR, digital twin (DT), tactile Internet, artificial intelligence (AI), and blockchain-based economy \cite{xu2022full}.

To create an immersive Metaverse experience, various technologies must be integrated and coordinated. In the following, we highlight the essential technologies for human-centric Metaverse design, but emerging technologies are not limited to this discussion as the Metaverse is continually evolving. Considering the left side of Fig. \ref{fig:human-centric-metaverse}  as an example, multiple technologies are used to create a virtual avatar, including wireless sensors, sensor fusion, interpretation with machine learning, 3D projection from data, and 3D view synthesis. Machine learning/deep learning algorithms can be utilized to fuse the data collected from the sensors effectively \cite{winkler2022questsim, yi2022physical}. These learning algorithms can further capture user data patterns, e.g., body shape, behaviours, poses, and expressions, and then prepare the data to be projected into the virtual environment. Finally, the virtual avatar is placed into the virtual scene (with VR) or mixed environment (with XR). The user's experience in the VR/XR environment can be enhanced by optimizing the view, angle, resolution, and interaction within the scene \cite{lee2021all}. Extra haptic feedback can also be utilized to generate realistic feelings about the virtual environment \cite{wu2017evaluation, Wu2016}. Other necessary technologies include intelligent sensing, data compression, edge computing, and wireless multiple access to reduce latency and improve reliability \cite{xu2022full}.

Beyond gaming, governments and tech companies seek a presence in the Metaverse. Decentraland lets users create, explore, and interact with 3D virtual worlds owned and controlled by themselves \cite{decentraland}. Users buy virtual land as NFTs via MANA cryptocurrency, which uses the Ethereum blockchain. NVIDIA's Omniverse introduces a computing platform for creating Metaverse applications such as 3D scene generation, art creation, and robotic control with supportive generative AI and physic-based simulation engines \cite{omniverse}. Microsoft's Mesh brings a new toolset for users to create custom workplaces and tools harmonized with other applications in Microsoft's ecosystem, such as Teams \cite{msmesh}. Other Metaverse apps focus on healthcare and education, such as Xirang \cite{xirang} and Telemedicine \cite{telemedicine}.

Besides using conventional sensing and data collection techniques, integrating human physiological and psychological information is crucial for developing human-centric Metaverse applications. As motivated by the fact that the brain signals are encoded with rich information about human activities, in the following, we discuss the details of BCI  technology and how BCI can offer multimodal, low latency, and high fidelity metrics for Metaverse user behavior \cite{Blankertz2006}.

\subsection{Brain-Computer Interface: Sensor Technology}

\begin{figure}[t]
    \centering
    \includegraphics[width=0.8\linewidth]{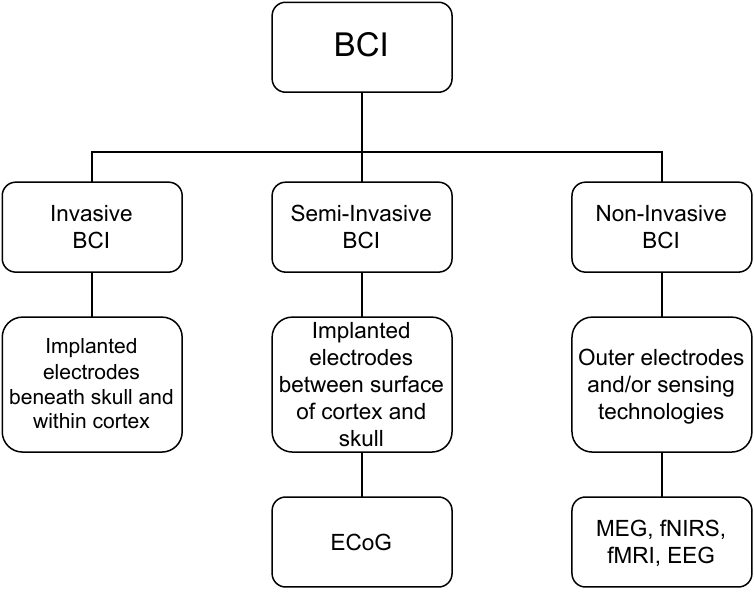}
    \caption{Three types of BCI devices/sensors: invasive BCI, semi-invasive BCI, and non-invasive BCI.}
    \label{fig:bci-types}
\end{figure}

The human brain is the most complex and adaptive organ within the human body \cite{Ungerleider1994}. It is the control center of human intelligence, sensory perception, and motor functions. Herculano-Houzel \cite{HerculanoHouzel2012} estimated that the central brain might contain around 86 billion neurons. Each neuron is a node along trillions of neural pathways within the brain. Each neural pathway passes neuroelectric signals (neurotransmission) around the brain, forming a system that enables the brain to function by communicating with the nervous system. A broad definition of Brain-Computer Interface (BCI) is any device that measures, analyzes, and interprets the brain signal (neural pathways) and then relays information to a machine to respond. The story of BCI begins with a discovery made by the British physicist Richard Canton \cite{haas_2003}. In 1875, Canton discovered the existence of electrical signals in the brain of animals. This discovery paved the way for the pursuit of electrically mapping brain signals and a better understanding of human neurophysiology. Four decades later, a psychiatrist named Han Berger invented the first measurement device, allowing humans to measure the brain's electrical activity for the first time \cite{haas_2003}. Berger created the tool and discovered the first neural oscillation frequency, the 8-12 Hz Berger (Alpha) wave. In modern times, researchers furthered these discoveries by developing various ways to measure the brain's neural signals, learning new neurophysiological behaviors, and building autonomous systems. BCI refers to technologies that can create communication pathways from the brain's activity to external devices, such as a computer or a machine \cite{Blankertz2006}. BCI devices/sensors are delivered in one of three forms, invasive, semi-invasive, and non-invasive, as illustrated in Fig.~\ref{fig:bci-types}. 

\textbf{Invasive BCI} devices characterise electrodes implanted (through surgery) beneath the skull and within the cortex (direct signal acquisition from the brain). Invasive BCI systems are not in a mature stage of development to be safely used as consumer devices. Invasive BCI require longer dedicated research with animal populations before engaging in low-sample-sized human research studies \cite{Waldert2016}. Invasive BCI is often employed in extreme cases (e.g. severe motor disability) where the patient's quality of life benefits from the BCI outweighs the risks \cite{Anitha2019}. Companies, such as Neuralink, are pursuing the goal of implantable BCI devices \cite{Fiani2021}.

\textbf{Semi-Invasive BCI} devices are sensors that are implanted (through surgery) between the cortex (on the surface) and the skull \cite{panoulas2010brain}. Semi-invasive BCI systems commonly use an array of Electrocorticography (ECoG) electrodes to map a specific brain region. The surgical procedures and implant durability for semi-invasive BCI typically carry lower short to long-term risks when compared to invasive BCIs. Semi-invasive electrodes offer a higher-quality signal; however, similar to invasive BCI, the surgical risks outweigh the benefits of the device. Due to the risks of invasive and semi-invasive devices, non-invasive systems are more popular as a low-risk and more viable product for researchers and consumers. With further research, reduction in surgical risks, and improved robustness of implants, invasive and semi-invasive BCI devices will have great potential to enhance the user experience within the Metaverse dramatically.

\begin{figure}[t]
\centering
\includegraphics[width=1.0\linewidth]{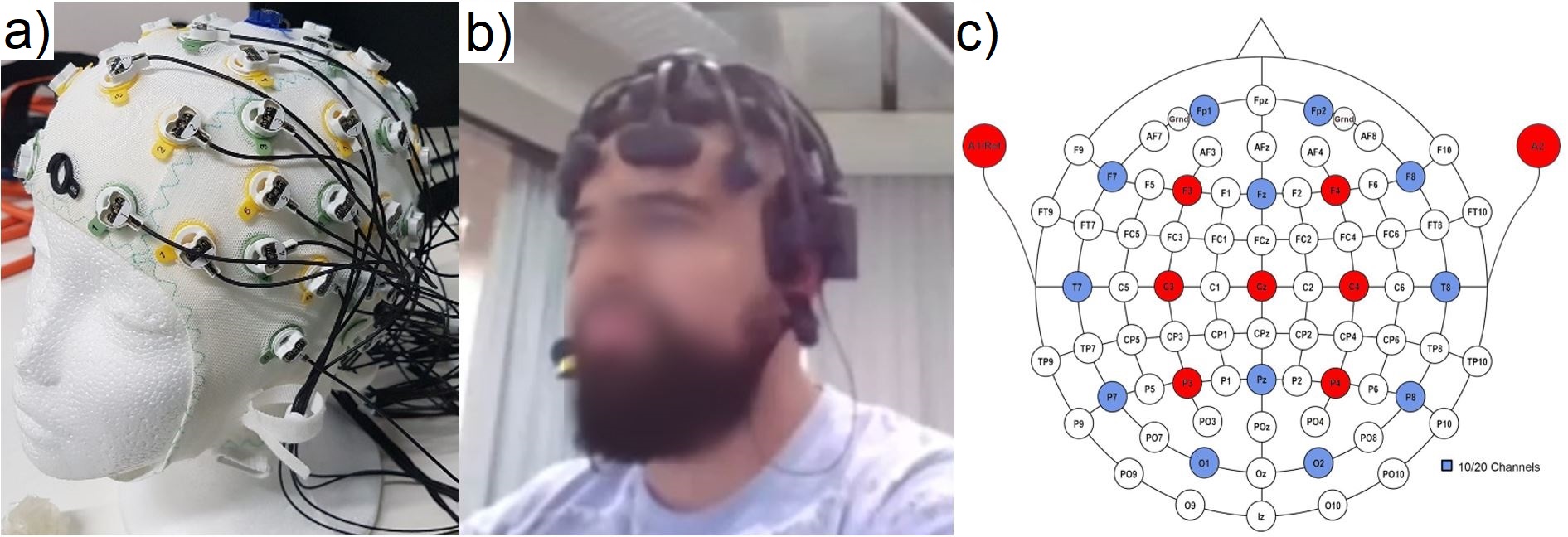}
\caption{This figure presents the two types of non-invasive EEG-based BCI devices available on the market: (a) the Brain Products' actiCAP active 64-channel wet EEG electrodes system; (b) the Cognionics Quick-20 dry EEG electrodes system; and (c) the layout for both the 10/20 EEG channel layouts used in both systems. Pictures were taken from the Computational Intelligence and Brain-Computer Interface (CIBCI) lab at the University of Technology Sydney (UTS), Australia.}
\label{fig:BCI_Electrodes}
\end{figure}
 
 \textbf{Non-Invasive BCI} devices encompass multiple technologies that can detect neurophysiological behaviours without any implanted electrodes; this includes technologies such as Magnetoencephalography (MEG), Functional Near-Infrared Spectroscopy (fNIRS), Functional Magnetic Resonance Imaging (fMRI), and Electroencephalography (EEG). Certain non-invasive BCI systems, such as fMRI and MEG, lack portability due to equipment size or complexity. These types of systems are not feasible as wearable devices for Metaverse users. EEG and fNIRS-based BCI systems are the primary feasible solution for a portable, wearable, and accurate device that can be used in a general consumer capacity \cite{Wolpaw2000}. Typically, EEG devices use wearable scalp electrodes (see Fig.~\ref{fig:BCI_Electrodes}) with a highly conductive material to measure the voltage ($\mu$V) fluctuations on the wearer's scalp \cite{Kerous2017}. On the other hand, fNIRS electrodes utilize near-infrared spectroscopy to discern cortex neural activity \cite{Wilcox2015}. Certain systems offer paired EEG-fNIR electrodes within one system \cite{Uchitel2021}. These types of electrodes will measure neural signals, which are then amplified and digitized for analysis. The resulting signal may contain multiple components (depending on electrode placement), such as eye blinking, muscle movement, movement artifacts (from displaced channels), and other electrical activity. Most importantly, the signal will contain information on the brain's electrical activity \cite{Minguillon2017}. Through extensive repeated measure research and machine learning, common neurophysiological behaviors can be classified and used for various research applications, e.g., military, rehabilitation gaming, medicine, mental health, robotics and automation, and public services \cite{Marshall}. Therefore, wearable non-invasive EEG/fNIRs BCI devices are most suitable for researchers and consumers exploring the Metaverse.

EEG devices typically consist of two types, wet and dry (see Fig.~\ref{fig:BCI_Electrodes}) of electrode systems \cite{Gordo2014}. Wet electrodes refer to any electrode system that requires conductive gel or saline fluid to improve the contact connectivity between the electrode and the wearer's scalp. In contrast, dry electrodes leverage optimized electrode shapes (often hair comb-like) to contact the scalp without needing gel/fluids. When comparing the two types, wet electrodes offer a better signal quality (less noise from impedance and external sources) but require preparation (applying gel/fluid) and a limited operation time due to the drying of the gel/fluid. Dry electrodes are generally larger than wet electrodes, limiting the total possible electrodes placed on the scalp, the overall signal quality (large electrodes are more susceptible to movement), and the user comfort when wearing the device \cite{Hairston2014}. Both types of EEG systems are limited by movement noise, device usability (user comfort and real-world practicality), and neurodiversity \cite{Minguillon2017}. Unlike EEG, fNIRs do not require conductive gel as the sensors primarily use light \cite{cay2017design}. The key drawback to consumer fNIRS devices is the requirement of paired sensors (a source and detector) to measure neural signals. A 64-channel system would require 128 fNIRS sensors compared to the 64 EEG electrodes. The current size of the fNIRS sensors makes the system unideal for consumer use, as dry EEG electrodes could achieve a similar result with fewer sensors. Therefore, given the real-world feasibility factors, EEG dry electrode BCI systems are currently ideal for Metaverse users. It is likely that with further improvement to signal processing techniques, machine learning algorithms, and hardware design, we will find that dry electrode EEG with fNIRS BCI devices will become the next popular consumer device.

 \begin{figure}[t]
\centering
\includegraphics[width=1.0\linewidth]{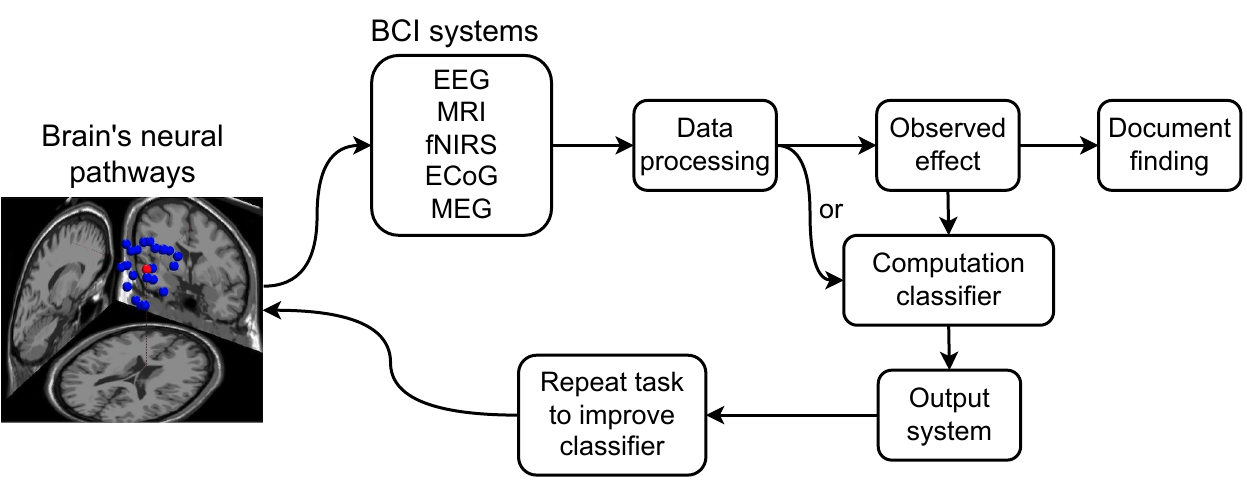}
\caption{An overview of the pipeline of BCI systems. The figure illustrates the acquisition of a signal through various types of BCI systems. Once a signal has been acquired, information can be extracted for observational purposes (monitoring or measuring a state) or classified into a specific behaviour (detecting intentions or tasks).}
\label{fig:BCI_Sig}
\end{figure}

Signal processing is another important aspect of BCI devices. Fig.~\ref{fig:BCI_Sig} outlines the typical workflow of BCI-related research \cite{Torres2020}. The two principal methodologies further enhance our knowledge of the neurophysiology of the brain to enhance BCI systems through signal feature recognition (built through observational information). The other is to develop real-time systems through classifiers (machine learning or AI). BCI research has expanded to many interdisciplinary research fields, studying behaviors such as cognitive states, emotional responses, pathology (neurology), mental health, pedagogy, ergonomics, and many other fields \cite{Mudgal2020}. The current challenge for BCI systems is to develop a real-time ``plug and play" system for consumer use \cite{Saha2021}.

\begin{figure*}[t]
    \centering
    \includegraphics[width=0.8\linewidth]{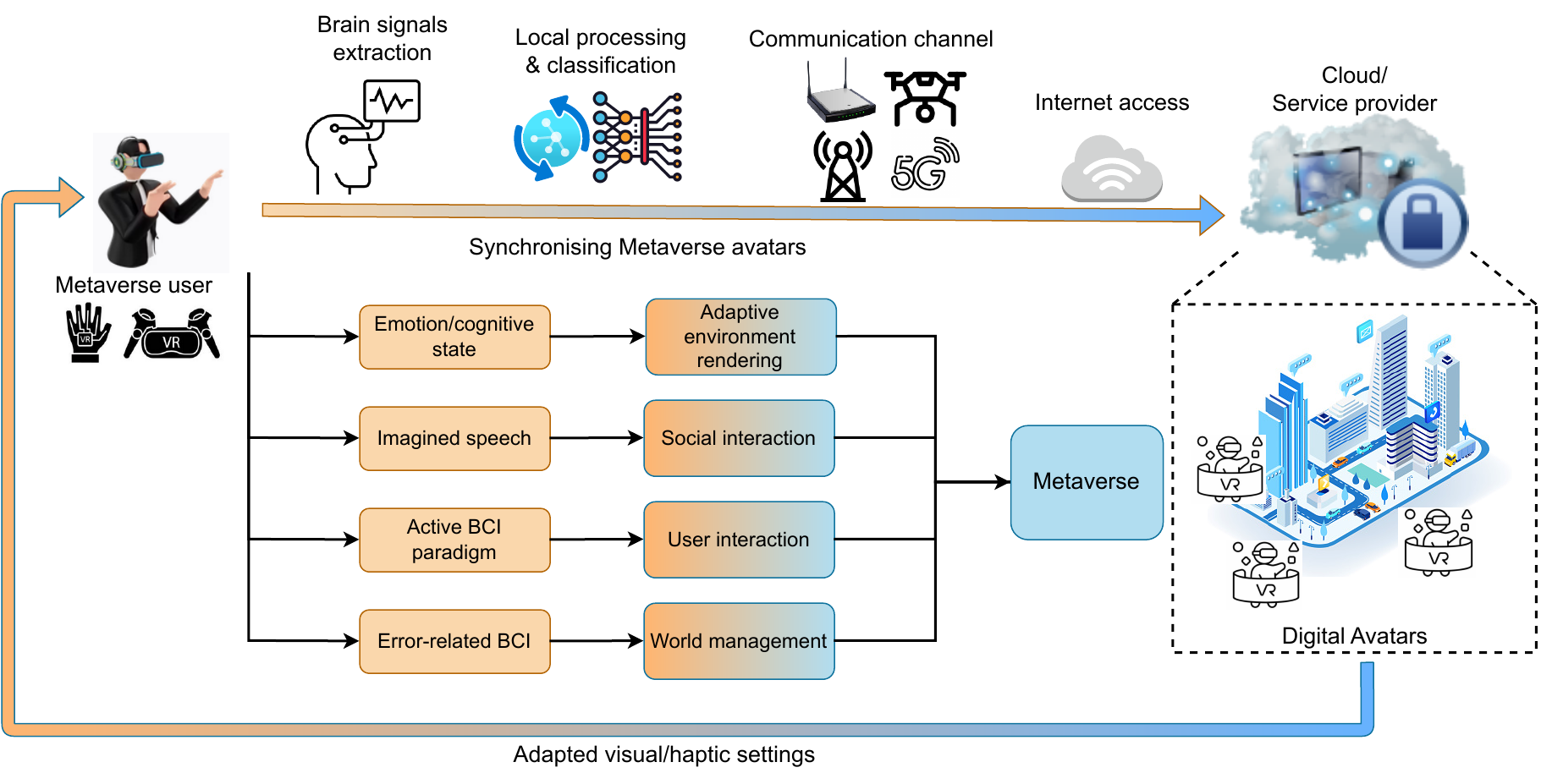}
    \caption{An illustration of the integration of BCI with the Metaverse. Through BCI, the Metaverse user's brain signals can be extracted, processed, and communicated into the Metaverse. The figure outlines the types of information that can be obtained from the BCI device and integrated into the Metaverse.}
    \label{fig:background}
\end{figure*}

\subsection{Deployment of BCI within the Metaverse}
\begin{figure*}[t]
    \centering
    \includegraphics[width=0.7\linewidth]{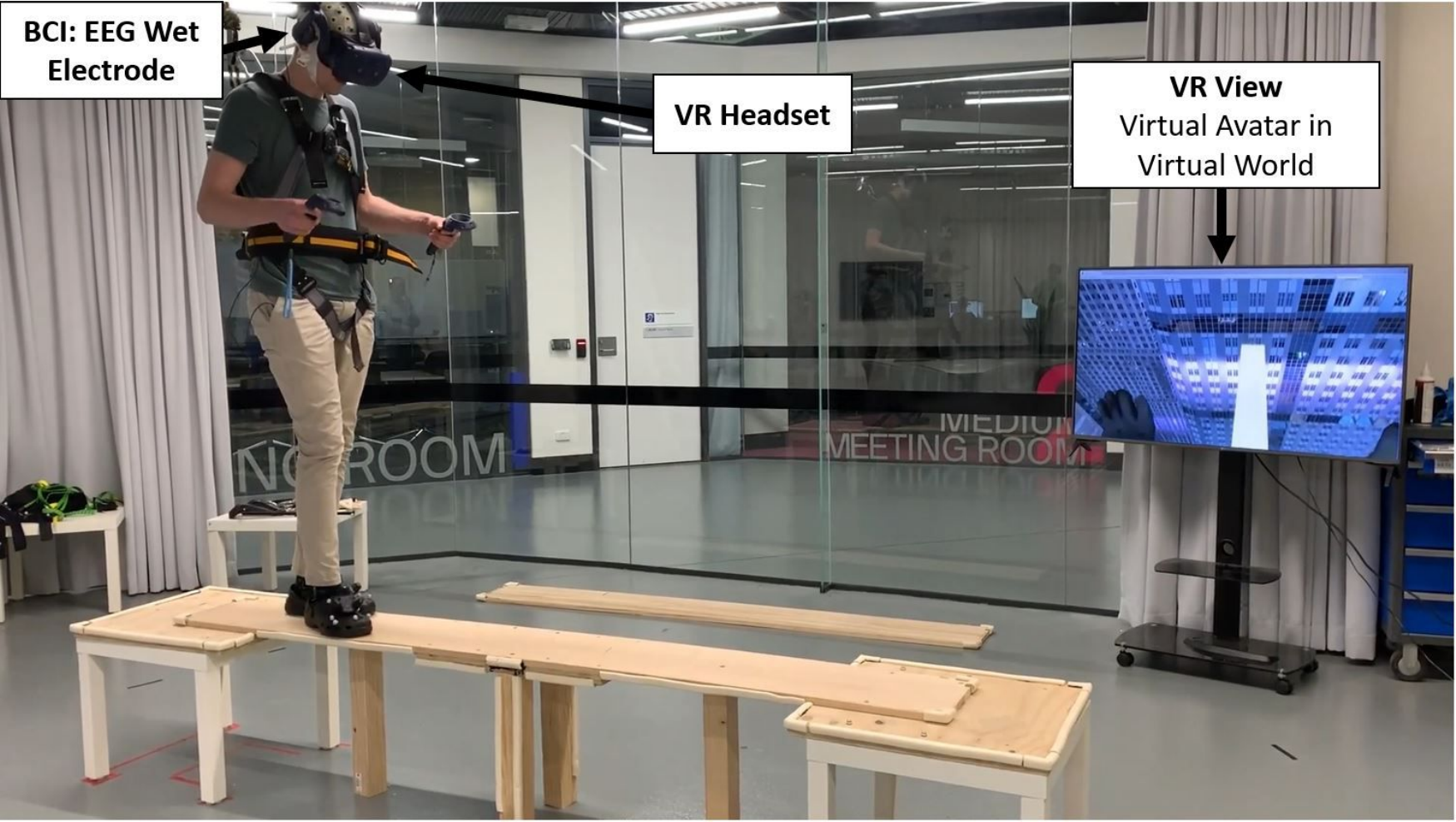}
    \caption{An illustration of a VR user using a VR headset (HTC Vive Pro) with a BCI sensor cap (64 channel EEG, Liveamp system) worn under the headset. The user is experiencing a mixed reality environment where they are physically (through the platform) and virtually (through VR) elevated. The picture was taken from the Computational Intelligence and Brain-Computer Interface (CIBCI) lab at the University of Technology Sydney (UTS), Australia.}
    \label{fig:VR_BCI}
\end{figure*}

In Fig.~\ref{fig:background}, we illustrate a BCI-enabled Metaverse in which BCI plays a vital role as an interface to create adaptive virtual environments and intelligent avatars, supported by other technologies such as a digital twin and real-time communications. As illustrated in Fig.~\ref{fig:background}, a BCI-enabled Metaverse is a human-centered approach in which BCI and VR technologies can co-exist and cooperate in a closed loop. Within the conventional BCI research field, there are many examples of VR-BCI integration \cite{zhu2021,Zhu2023,tirado2021,Singh2021} through using a traditional wet electrode EEG cap (see Fig.~\ref{fig:bci-types}(a) and Fig.~\ref{fig:VR_BCI}) under XR/VR device. This method of VR-BCI integration is viable in a research context because of the importance of signal quality and spatial resolution from the BCI operating in a controlled environment. This VR-BCI set-up would not be feasible for a real-world consumer because the wet sensor would only provide limited usage as the gel would rapidly dry out. A commercially available option to enable VR-BCI is to use dry electrodes integrated into the VR/XR device, such as the Galea VR HMD \cite{Bernal2022}. A dry electrode system will enable a portable system with lower signal quality and spatial resolution.

The basic operation of the BCI-enabled Metavese may include the following steps. The VR-BCI interface extracts the users' brain signals and processes the signals locally or remotely at a computing unit, e.g., a remote server. Brain signals extraction, processing, and classification are enabling processes for creating human-like digital avatars with unique characteristics of the users, e.g., emotional state, visual stimulus, and behaviors, from the human brain signals. The communication channels, such as wired and wireless channels, further enhance the scalability of the Metaverse system. The Internet-connected computing unit will update and synchronize the information into the Metaverse. 
By monitoring the brain activities of the users, the Metaverse platform can analyze or predict the users' behaviors, attention, or emotional states and adjust VR settings to be transmitted back to the users. As such, the service provider can actively provide customized and personalized Metaverse applications for the users. Note that other users in the Metaverse also contribute to the dynamics of the above process. In addition, other supporting technologies, such as digital twins, integrated VR-BCI devices, and real-time communications, can facilitate and improve the completeness of the system. BCI can be directly used to measure Emotion/Cognitive states of the Metaverse user and facilitate social, active, and passive interaction within the Metaverse.

\section{BCI-Enhanced User Immersion}
\label{sec:bci-enhanced-embodiment}
\begin{figure*}[t]
    \centering
    \includegraphics[width=0.8\linewidth]{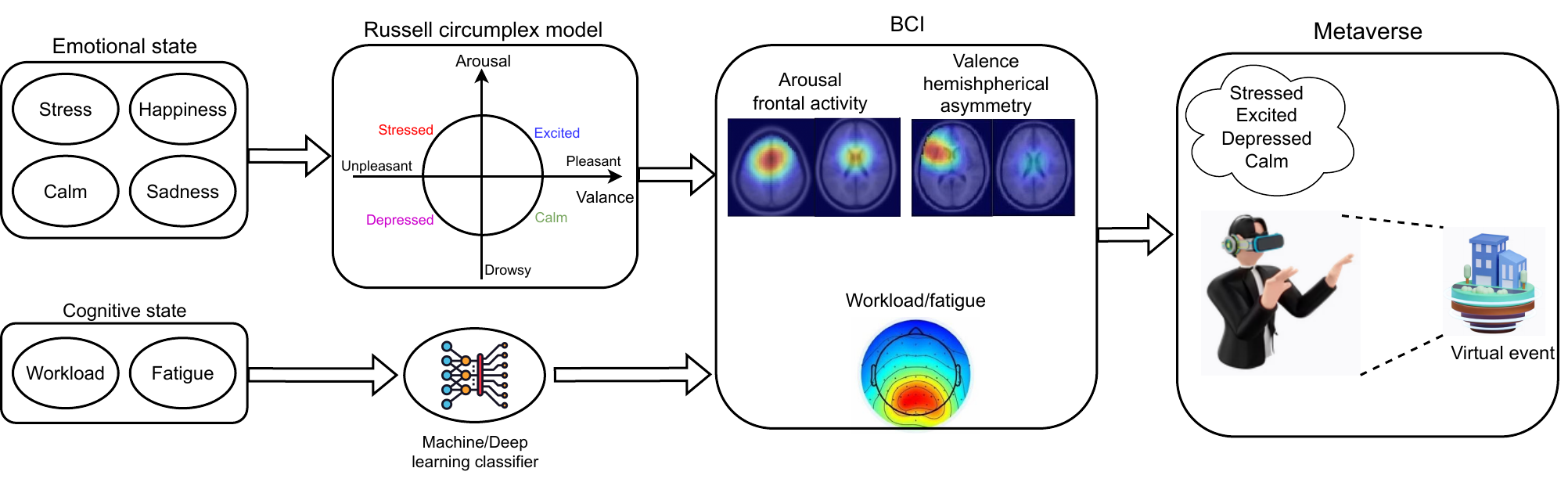}
    \caption{An illustration of the emotional and cognitive state measurements using BCI. These types of information can be utilised in the Metaverse to provide status indicators and improve the immersion of Metaverse users.} 
    \label{fig:Emotional_Cognitive_State}
\end{figure*}

Immersion is the degree of realism for the congruence (between the real and virtual worlds) of the sensory input and motor output. Immersion plays an essential role in the Metaverse user experience. A fully immersed Metaverse user can seamlessly transition between the physical world and the Metaverse. Therefore, the modulation of the immersion levels of the environment can directly affect the believability and acceptance of the Metaverse. BCI can improve the immersion of the Metaverse user by altering the rendering of the virtual environment based on the user's emotional and cognitive state. Previous works explored this concept \cite{cho2014bci} using passive BCI techniques to adaptively change the game environment's lighting and Field of View (FOV), thus enhancing the user's immersion. Passive BCI refers to using BCI to detect and measure changes in a user's unintentional cognitive and emotional state \cite{arico2017passive}. In this section, we explore two potential methods of using BCI to enhance the user's immersion in the Metaverse. Specifically in Section \ref{sec:EmotionalCog}, we explore the user's emotional (including happiness, sadness, stress, calmness, anxiety, and uneasiness) and cognitive state (mental workload, fatigue, and attention) as shown in Fig.~\ref{fig:Emotional_Cognitive_State}. After that, in Section \ref{sec:Error}, we present the potential of using anomalous and error-related neurological behaviors to enhance immersion by correcting anomalies within the Metaverse.

\subsection{BCI Emotional and Cognitive State Recognition Applied to the Metaverse}
\label{sec:EmotionalCog}
\begin{figure}
\centering
\includegraphics[width=0.9\linewidth]{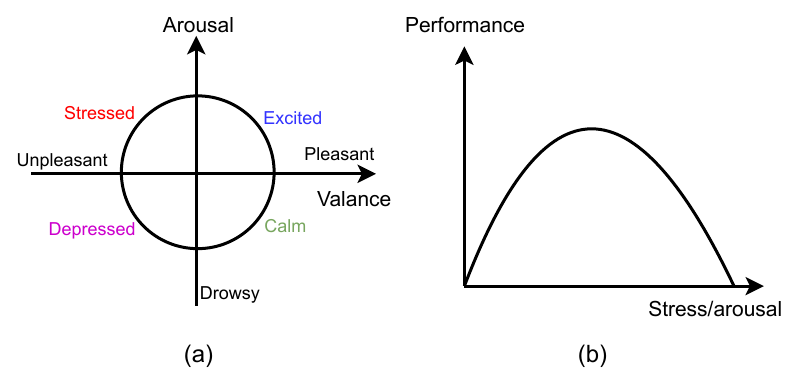}
\caption{\title{The Russell Circumplex Model of Affects and the Yorkes-Dodson Law} This figure presents (a) the Russell Circumplex model of Affects, which is used to measure emotional states on a spectrum between arousal and valence; and (b) the Yorkes-Dodson Law dictates the relationship between emotional arousal/stress and cognitive performance. 
\label{fig:Emotional_Models}}
\end{figure}

A person's emotional state is commonly quantified by a scale of valance (pleasant to unpleasant) and arousal (alertness to drowsy). This mode of mapping a person's emotional spectrum is known as the Russell Circumplex model of affects \cite{Russell1980}. Fig.~\ref{fig:Emotional_Models}(a) presents the Russell Circumplex models and shows the spectrum of emotional states quantified by the arousal and valance level. This understanding was furthered by the Yorkes-Dodson law \cite{Broadhurst1957} (see Fig.~\ref{fig:Emotional_Models}(b)) that found a direct correlation between human emotional arousal and cognitive performance. Therefore, the ability to quantify and measure human emotional states can play an essential role in understanding an individual cognitive state. The BCI classification of emotional states involves extracting specific EEG features for arousal and valance from the EEG signal for a machine learning classifier to detect \cite{Nafjan2017}. Arousal is detected through the changes in the brainwaves in the brain's frontal region. Brainwaves are common oscillations in the brain's electrical activity that correlate to various neural activities. The brainwaves are broken down into the Gamma ($>$35 Hz), Beta (12-35 Hz), Alpha (8-12 Hz), Theta (4-8 Hz), and Delta (0.5-4 Hz) wave. Studies \cite{Bos2007,jenke2014feature,Stenberg1992} found a strong correlation between an individual's arousal level and the frontal Beta, Alpha, and Theta power. The ratio between Beta and Alpha activity is commonly used to measure arousal level. Typically, the Beta brainwave would indicate an active mental state. Conversely, the Alpha brainwave suggests a relaxed and restful state. Therefore, heightened arousal can be measured by a frontal region increase in Beta power and a decrease in Alpha power. An individual's valence level is measured through the brain's hemispherical symmetry/asymmetry. Hemispherical symmetry refers to an equal/similar activation (neuron firing) state between the brain's left and right cortex. Hemispherical asymmetry occurs when one cortex has significantly more activity than the other. Studies \cite{Prete2019,Gainotti2018,Mneimne2010} showed that valence correlates to the degree of hemispherical symmetry with a strong hemispherical asymmetry exhibited when in a negative valance (unpleasant) state. Works by \cite{Atkinson2016} and \cite{He2020} used BCI and machine learning to classify the dimensions of an individual's arousal and valence levels, which indicate their emotional state.

A person's cognitive or mental state refers to their mental well-being and the ability to think or process information. A Metaverse user's mental and cognitive state can significantly impact their experience within the Metaverse. A high workload (complex or challenging to navigate environment) or sensory-loaded (high noise or colour intensive) environment can trigger negative mental states, reducing the user's immersive in Metaverse. Factors such as the current emotional state, experience of mental workload, fatigue level, and attention level can directly affect a person's cognitive state. Like emotional states, cognitive state features are extracted by evaluating the brainwaves of specific brain regions. The theta activity in the frontal cortex often determines mental workload. Studies by \cite{so2017evaluation} and \cite{Berka2007} asserted that as the difficulty of a task (higher workload) increases, the theta activity in the frontal cortex will increase.

Interestingly, the inclusion of multimodal data sources such as cardiovascular (changes in heart rate) and pupillary activity (changes in pupil dilation) can improve the accuracy of the workload classification \cite{John2022}. Mental fatigue resulted in the increase in theta and alpha activity and the decrease of beta activity in the frontal cortex\cite{Jap2009}. Studies on attention discovered that a distracted (unfocused) individual would exhibit a decrease in beta power in the frontal region, an increased theta and delta power in the central region, and a decrease in alpha power in the parietal region \cite{Lin2022,Klimesch1998}. 

\begin{figure*}[t] 
\centering
\includegraphics[width=0.55\linewidth]{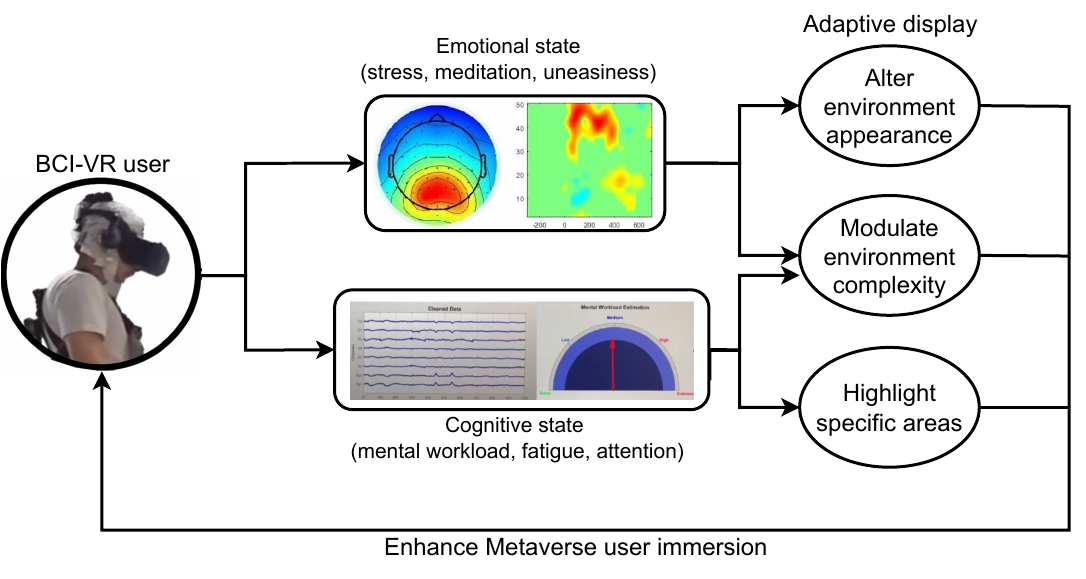}
\caption{\title{An illustration of passive BCI enhancing user immersion.} This figure shows using passive BCI to detect the Metaverse user's emotional and cognitive state. Then using, the measured information to modulate different factors of the Metaverse display to enhance the user's immersion.  The picture was taken from the Computational Intelligence and Brain-Computer Interface (CIBCI) lab at the University of Technology Sydney (UTS), Australia.  
\label{fig:BCI_Adaptive}} 
\end{figure*}

Fig.~\ref{fig:BCI_Adaptive} depicts using passive BCI to create an adaptive Metaverse display to enhance the user's immersion. Using passive BCI to gauge a user's emotional and cognitive state is a well-researched area with multiple reliable classifiers to enable the technology. When introduced to the Metaverse, passive BCI can dynamically adjust the user's surrounding environment and render display to improve the user's immersion. An example of this was the adaptive virtual reality environment by \cite{Sergi2019}. By measuring the VR user's emotions, the system created a feedback loop that used the virtual environment to modulate the user's emotional state. Similarly, the works by \cite{Wozniak2021} and \cite{cho2014bci} explored limiting and adjusting an environment's complexity to improve the user's cognitive state. These works would use adjustable lighting and fog to moderate the amount of the visible virtual environment to reduce the user's workload and visual fatigue. These practices could be applied to the Metaverse through an integrated BCI VR device to create a feedback system that adaptively adjusts the rendered environment. This solution is very close to realization with the recent innovations such as the workload measurement integration in the HP G2 Reverb VR HMD \cite{Omnicept} and the dry electrode BCI integrated Galea VR HMD \cite{Bernal2022}. 

\subsection{Anomalous and Error-related Behaviours to Improve User Immersion}
\label{sec:Error}
Another unique functionality of passive BCI is the ability to detect potential adverse events before consciously recognising the event. Adverse and anomalous events can hinder user immersion by creating a disassociation between the expected real-world and the Metaverse. Examples of this could be events such as the onset of VR sickness, loss of balance/falling, or environmental errors. The real-time detection of these events allows the implementation of safety and preventative measures to improve the user's longevity within the Metaverse. Through extensive studies, each adverse event's unique EEG signal features can be reliably extracted and classified.

VR sickness refers to the sensation and experience of symptoms such as headaches, nausea, vomiting, drowsiness, and disorientation when using VR \cite{tirado2019}. This is relevant for Metaverse applications, where users often spend prolonged periods in virtual environments. When VR sickness occurs, users often disengage from the virtual environment to alleviate the symptoms. In extreme cases, it may result in fainting, falling (due to loss of balance), or severe nausea. These types of adverse events will negatively affect the user's sense of immersion and reduce the longevity of the Metaverse user to stay within the Metaverse. Certain VR Studies \cite{tirado2020,Chen2010,Tirado2023} have successfully used EEG signals to classify and detect when a person is experiencing VR/motion sickness. They found a significant correlation between VR/motion sickness and theta and delta bands activity within the occipital lobe (attributed to the sensory perception of motion). They also observed a decrease in alpha activity in the parietal and motor regions. The loss of balance or falling is another significant risk for VR Metaverse users \cite{tirado2021}. Studies by \cite{sipp2013} and \cite{Annese2016} show that the beta and theta band activity in the parietal/motor cortex is closely related to losing balance and falling. These VR sickness and falling indicators can be trained through an AI classifier to detect anomalous events in real time during a Metaverse experience. The ability to effectively detect VR sickness and falling can allow the implementation of preventative techniques such as reducing the motion of the virtual environment and turning on VR see-through mode \cite{tirado2021}. 

\begin{figure}
    \centering
    \includegraphics[width=1.0\linewidth]{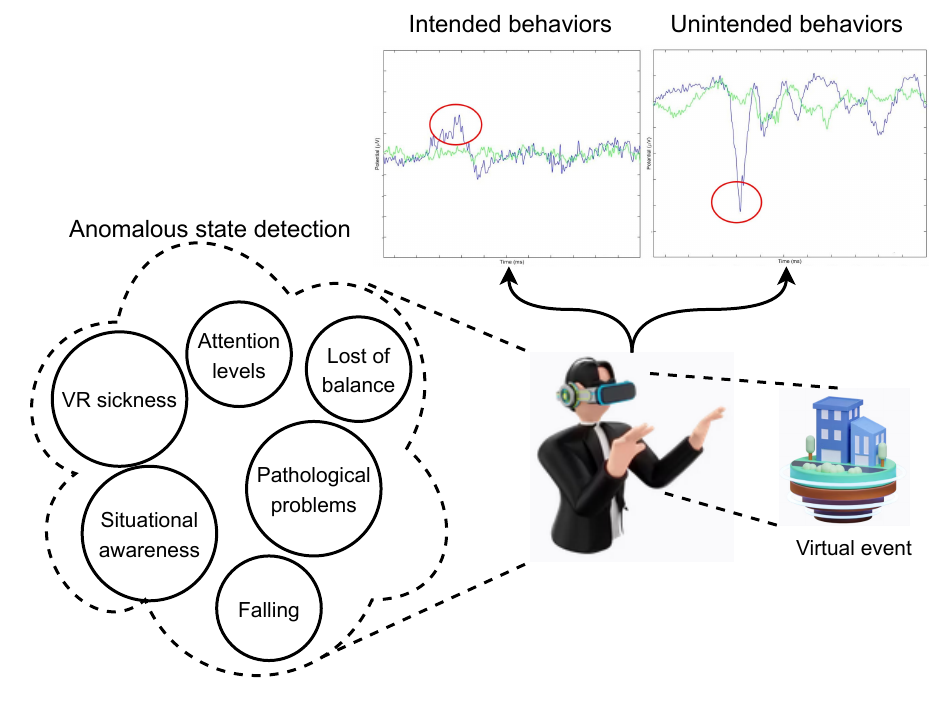}
    \caption{This picture illustrated the use of BCI to detect anomalous states and error-related behaviors. Using the BCI signal (e.g. ERN), the system can detect and correct adverse events, such as when the user is about to fall due to VR sickness.}
    \label{fig:Error_Anomaly}
\end{figure}

In the continuity of the Metaverse, errors and visual bugs are inevitable factors that will appear within the virtual environment. Erroneous artifacts or system glitches can hinder a user's immersion.  Therefore, it is essential to have a system in place to detect and correct these errors. One proposed method that BCI could solve this issue through error-related negativity (ERN) detection (see Fig.~\ref{fig:Error_Anomaly}). ERN is a signal response that occurs when a person observes incongruent or erroneous stimuli within a task or environment \cite{Yeung2004}. ERN is characterized by a negative potential around 50-250 ms after the error \cite{singh2018}. Previous studies\cite{Singh2021,Hakim2020} have successfully classified the ERN response as an error correction method within a VR environment. Due to the simplicity of ERN, it can be reliably implemented to detect potential errors that Metaverse users may observe. This solution would enable a more efficient (compared to manual user reporting) method of detecting and correcting potential errors within the Metaverse environment. 

These BCI solutions can improve the safety and longevity of a Metaverse user. By incorporating the outlined BCI techniques, the Metaverse system can become more reactive to adverse events and use an appropriate strategy to prevent or correct the problem. There are two critical challenges to the implementation of this system. The first challenge is ensuring the ability to accurately detect the onset of the adverse event before the occurrence or conscious recognition of the event. The system would not be valid if it cannot prevent the adverse event from occurring. The second challenge is the explore practical strategies for preventative and corrective methods. The current methods of prevention or correction involve removing the user from the virtual environment and breaking their immersion. Better methods are required that do not require the removal of the user from the Metaverse.

\section{BCI-Enhanced Metaverse User Interactions}
\label{sec:bci-enabled-metaverse}
An important aspect of the Metaverse is to deliver a platform to facilitate meaningful user interaction. A Metaverse user must be able to interact with the environment (pick up objects and locomotion) and socially with other Metaverse users. Active BCI offers the potential to generate more intuitive modes of user interaction within the Metaverse. In contrast to passive BCI, active BCI refers to the user of BCI performing specific tasks through intentional, conscious thought. This section will explore the ways BCI can be used for users to interact with their environment (Section \ref{sec:ActiveBCI}) and the potential of using BCI for social interaction through brain-to-text (Section \ref{sec:Social}).

\subsection{Decoding Thoughts and Intentions Using BCI to Improve Metaverse Interactions}
\label{sec:ActiveBCI}

\begin{figure*}
\centering
\includegraphics[width=0.55\textwidth]{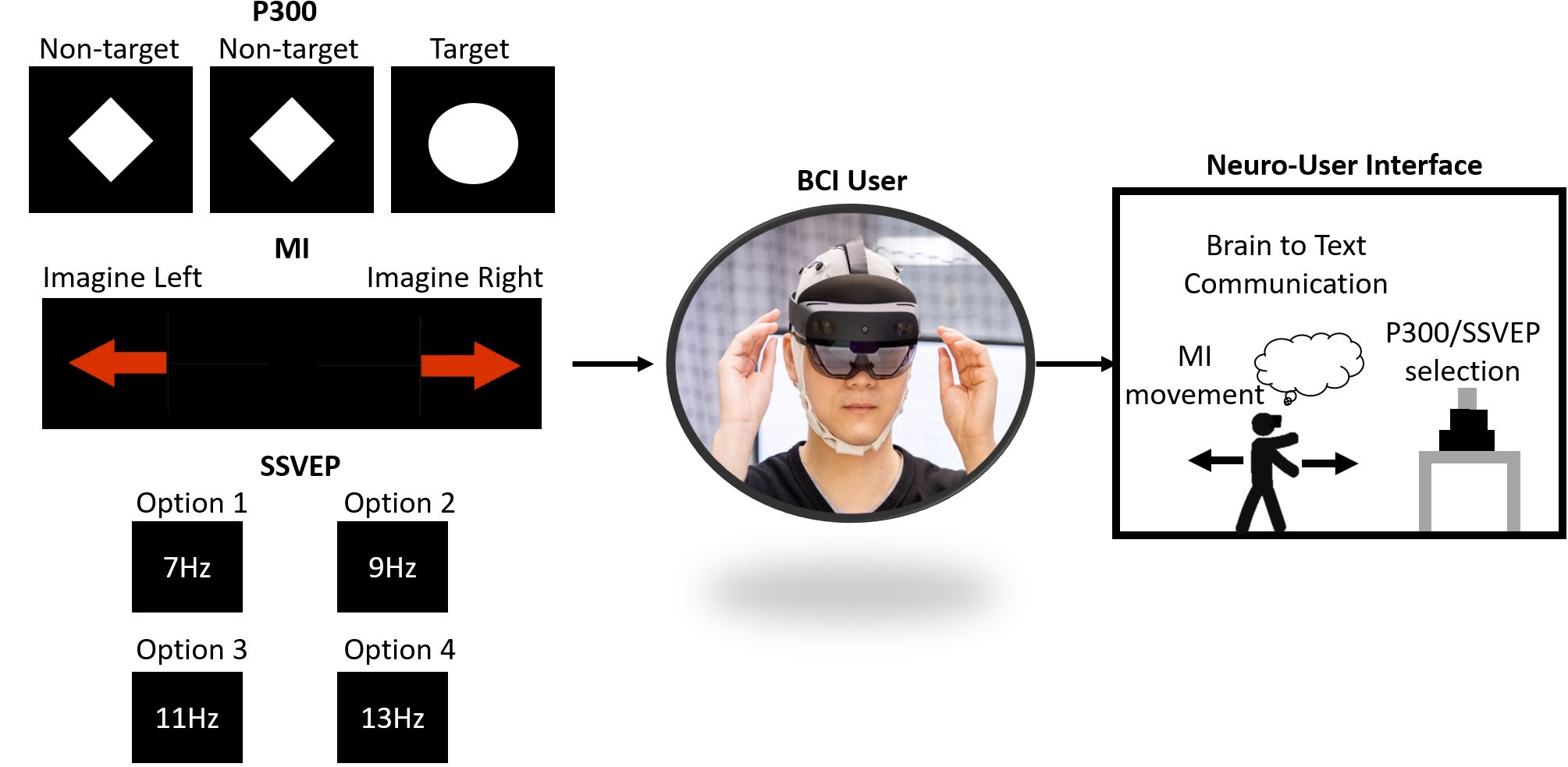}
\caption{\title{An illustration of how active BCI can enhance the Metaverse user interaction.} This figure presents the common BCI paradigms used for designing user interfaces and the potential of using brain-to-text for social interactions.
\label{fig:active_BCI}}
\end{figure*}

Understanding human thoughts and intentions is a commonly sought-after goal in the BCI research field \cite{Hinterberger2003}. Traditional systems rely on tactile manipulators, such as controllers, buttons, joysticks, levers, and keyboards, to allow users to convey their intentions to a system \cite{BalagtasFernandez2009}. Other works explored voice recognition, gesture control, and AI to develop more intuitive methods of understanding user intention \cite{Nehaniv2005}. BCI offers the potential for direct translation between human thought and intention, which could result in an intuitive and responsive system. The underlying challenge in understanding human intention is the complex multilevelled nature of the human mind \cite{Blakemore2001}. Human intention ranges from low-level cognitive decisions based on sensory perception (reacting to events, bodily movements, or simple choices) to complex high-level decisions requiring observation, planning, mental stimulation, and multistep execution. Based on this challenge, researchers have designed reliable active BCI paradigms to capture specific behaviors that are exhibited across the human population. When designing an active BCI system, one typically selects a reliable BCI paradigm to translate intentional thought into a classifiable EEG signal. We will highlight three of the most common paradigms used for active BCI control (as shown in Fig.~\ref{fig:active_BCI}); these are P300, Motor Imagery (MI), and Steady-State Visual Evoked Potential (SSVEP).

\textbf{P300:} The P300 wave is the oldest and potentially the most well-known EEG response out of the three paradigms. As described by \cite{Picton1992}, the P300 wave is a positive peak human event-related potential that occurs around 300ms after a `target' stimuli are perceived. This P300 peak can be observed across the brain's frontal, central, and parietal regions. The stimuli used for P300 paradigms can be both visual and auditory. Typically, P300 paradigms feature an oddball design where the user has a target and several non-target stimuli. A signal classifier can discern whether the user observes a target stimuli by detecting the positive peak amplitude. The P300 speller \cite{Krusienski2008} is a successful example of using a P300 wave to create a user interface. In this paradigm, the user will decide their target or intended letter to type; then, multiple visual letter stimuli will sequentially appear to the user. The classifier will detect the P300 response within the EEG signal and input the letter that appeared 300ms before the peak as the user's choice. Generally, P300 paradigms provide a reliable signal response for detection; the primary drawback is the speed of stimuli presentation (slow rate of input) and the dependency on user focus on the target stimuli (difficult to use in complex environments). 

\textbf{MI:} An MI paradigm utilizes the thought of left and right motor action to create a simple control paradigm \cite{Lotze2006}. MI involves the extensive training of a classifier that detects left and right motor actions (participants will clench their left or right hand) within the motor cortex. The classifier model relies on the hemispherical activations between the left and right-hand actions. Once sufficient training is complete, the user will train the classifier using the thought of left and right motor actions. Previous studies \cite{Crammond1997,Lotze2006,Decety1996} found that the mere thought or representation of a motor action can trigger a response in neurological pathways for actual motor action. This results in a reliable and detectable signal for left and right control without the need for 'real' motor action (not even tensing). MI's primary benefit is the lack of need for visual or auditory stimuli. However, MI requires extensive training individualized for each user and is more susceptible to noise if the user is mobile. 

\textbf{SSVEP}: The SSVEP paradigm is popular with multiple visual flashing stimuli that flicker at specific frequencies. SSVEP BCI studies \cite{Wu2008,Muller2008} show that when a person observes flickering/flashing visual stimuli, a synchronized frequency activity can be observed in the occipital region of the brain. In essence, if multiple flickering input control options are presented to the user, the frequency of the occipital region's activity can be used to determine which control option the user is focused on or targeting. SSVEP offers the advantage of not requiring extensive training while being a reliable paradigm for detection. The main drawback of the SSVEP paradigm is the argument that the stimuli require too much attention, as flickering visual stimuli will likely distract or block out the surrounding environment. 

Basic interactions and low-level intentions can be translated through active BCI paradigms such as P300, MI, and SSVEP. High-level decisions present are far more challenging to decode. Many works explored various areas of the hierarchy of executive decision-making \cite{Tidoni2017}. One example of a higher-level decision-making process is active spatial navigation \cite{Do2021}. The work investigated the use of BCI during active spatial navigation to understand the neurophysiological behaviors of a user when processing spatial information when navigating complex environments. The authors observed that the retrosplenial complex (RSC, a cortical region of the brain) theta activity correlates when a user engages in active spatial navigation (identifying spatial locations around the user). These biomarkers can provide an indication of whether a user is lost within a complex Metaverse environment. Other works explored the use of AI to build training models which observe specific behaviours/interactions of the user and attempt to decode the neurological behaviours (such as robotic systems interaction \cite{Zhang2020,Yue2021,Neuper2009}. In conjunction with other multimodal information and more advanced AI modelling, BCIs can be used to infer and decode many complex thoughts or intentions. These mechanisms can be applied to the Metaverse to understand the user better and facilitate a more intuitive user experience.

\subsection{Social Interaction Using Imagined Speech}
\label{sec:Social}

\begin{figure}[t]
    \centering
    \includegraphics[width=1.0\linewidth]{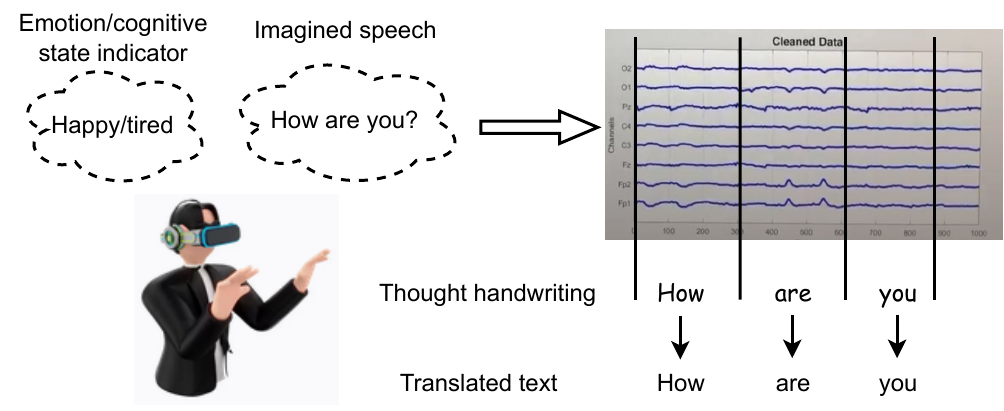}
    \caption{An illustration of using imagined speech enabled by BCI to perform thought-based social interactions. The figure depicts an example of how EEG data can be decoded into handwriting and then text to be used for imagined speech communication.}
    \label{fig:Social}
\end{figure}

The ability to input semantic information is an essential form of communication in human society. In the digital era, the keyboard has become a ubiquitous tool that enables a human to translate written language into a digital form that can be communicated between humans and computers. A key challenge in the Metaverse is to provide effective communication methods for social interaction. The straightforward approaches are to use a virtual keyboard \cite{Wu2016} or use a microphone for direct speech \cite{Maloney2020}. In the traditional line of thought, BCI can offer keyboard and letter selection solutions through P300 \cite{Krusienski2008}, MI \cite{Obermaier2003}, and SSVEP \cite{Hwang2012}. However, it could be argued that these methods would be inherently less efficient than traditional ones. We believe that the more significant application of BCI for communication is the potential of creating a new communication medium called `imagined speech'. By decoding human thoughts, semantic comprehension, and emotions, BCI could allow individuals to communicate through their thoughts alone \cite{WeiWang2011,Rabbani2019}. 

A recent innovation is the research by \cite{Willett2021}, which explores decoding brain activity into text. This offers the ability for the Metaverse user to interact socially by thought alone. Various works achieved this feat \cite{Herff2015,Willett2021,Herff2017} in which ECoG, an implanted electrode grid, provides spinal cord injury patients with the ability to generate text through thought. This technique decodes the brain's motor cortex region to interpret the thought or representation of the handwriting motor action (similar to MI). Then, the handwritten text is translated via deep learning into digitized signals. As suggested by \cite{lee2022toward}, this technology enables imagined speech communication between the Metaverse and real-world users. In principle, the finding from the ECoG results can be applied to EEG BCI devices. Fig.~\ref{fig:Social} illustrates an example of using EEG signals to classify imagined speech for social interactions. The critical ongoing challenge is translating this work from ECoG to EEG. The semi-invasive BCI system (ECoG) is a direct form of brain activity sensing with minimal noise. In contrast, a non-invasive EEG BCI system may produce significantly worse signal quality. Another challenge is that neurodiversity and the requirement of extensive participant training will hinder the realization of imagining speech technology. 

The outlined BCI paradigms offer the potential to build a more intuitive user interface and social interaction for Metaverse use. The ongoing challenge is to develop a reliably detected paradigm in an EEG signal, which requires minimal training, does not unnecessarily distract the user, and can be used by various users. These ongoing challenges indicate the need for further research and exploration into BCI to create a technology used for the Metaverse user.

\section{Human Digital Twin}
\label{sec:human-digital-twin}
Sociality is a key factor in the Metaverse \cite{ning2021survey}. The Metaverse must facilitate a social environment with continuity and a stable population of users. A challenge within the Metaverse is an unstable user base and the discontinuity of interaction when a Metaverse user exits the interfacing device (XR/AR/VR or a smartphone). This section examines the potential solution for this problem by developing a Human Digital Twin(HDT) in the Metaverse using BCI and Digital Twin (DT) technology. DT is a computer-based model that simulates and emulates physical entities, such as objects, humans, or human-related features, with their digital counterparts \cite{barricelli2019survey}. While DT has been applied in various areas, such as manufacturing, healthcare, and the Internet of Things (IoT), the concept of HDT remains largely unexplored. A few works \cite{barricelli2020human, shengli2021human} attempted to define HDT paradigms. Still, these primarily focus on using conventional IoT sensors to collect human information, similar to DT approaches in healthcare, such as fitness management \cite{martinez2019cardio} and disease diagnosis \cite{barricelli2020human}. To our knowledge, no work has considered using human brain signals to construct HDT avatars.

We propose the development of an HDT as the ultimate Metaverse BCI prosthesis (see Fig.~\ref{fig:hdt-model}). The HDT will create a stable population of Metaverse users that maintain continuity between the Metaverse and the real-world. Using real-time data metrics, such as EEG, smartphone, and smartwatch data, the HDT will behave in the Metaverse as an extension of the user in the real-world (when the user is not present in the Metaverse). The HDT will interact with the Metaverse, other Metaverse users, and other HDT. The user can replace the HDT by accessing and entering the Metaverse. In this section, we will outline the wearable technologies required to achieve HDT (Section \ref{sec:HDTBCI}), the potential applications of HDT within the Metaverse (Section \ref{subsec:ApplicationHDT}, and the potential challenges for the development of HDT in the Metaverse (Section \ref{subsec:real-time-commun}).

\subsection{Key Technologies to Enable the Human Digital within the Metaverse}
\label{sec:HDTBCI}
\begin{figure*}[t]
    \centering
    \includegraphics[width=0.7\linewidth]{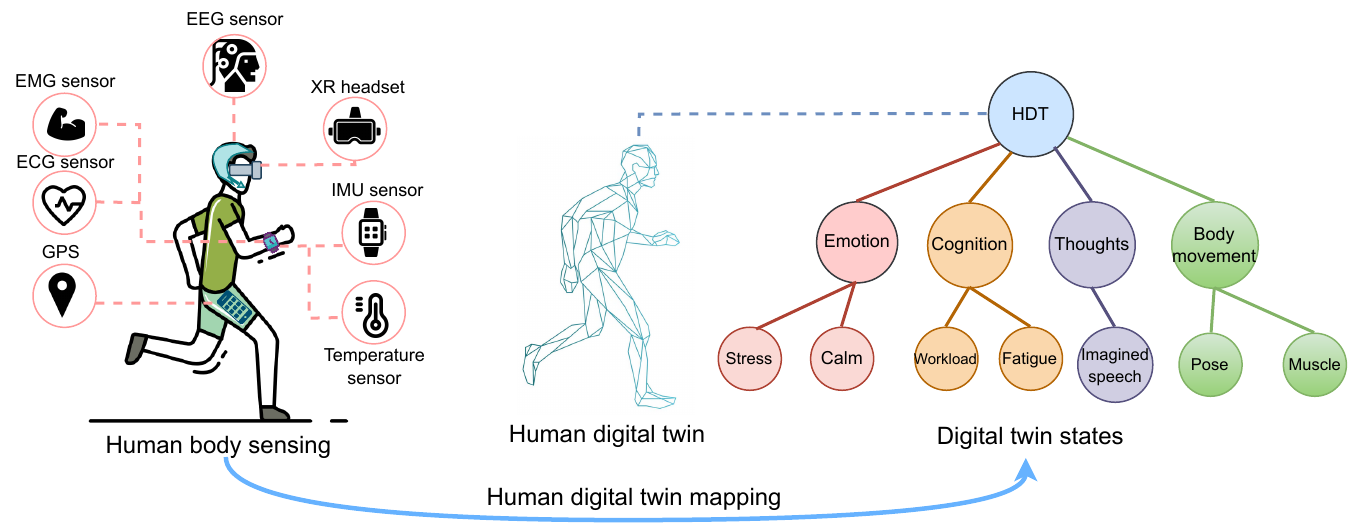}
    \caption{An outline of the key technologies and information involved in mapping the Human digital twin from human body sensing (with BCI and other supporting technologies). The graph model illustrates the human digital states containing real-time information of the human user.}
    \label{fig:hdt-creation}
\end{figure*}
Fig.~\ref{fig:hdt-creation} outlines the core components of creating the human digital twin. In this survey, we identify several technologies enabling HDT within the Metaverse. Technologies such as BCI, wearable biosensors (heart rate, muscle, IMU, and temperature), smartphones, and AI, can be integrated into the Metaverse to formulate the HDT to replicate life-like representation of the user's cognition, emotions, thoughts, and movements. 

\subsubsection{Wearable Brain-Computer Interface}
One clear advantage of using human brain signals to construct HDT avatars in the Metaverse is the potential reduction of the number of sensors and wearable devices required for users, leading to lower production costs and increased mobility and creativity. The brain signals contain a wealth of information about physical and mental health, such as the ability of EEG to complement electrocardiograms in predicting and diagnosing indicators of pathological perturbations, as demonstrated in brain-heart interaction studies \cite{lin2016delay}. As a result, the number of electrocardiogram sensors may be reduced or eliminated. Using BCI to control prosthetic devices, as described in \cite{katyal2014collaborative}, opens the door to potential BCI applications such as performing activities of daily living. Other BCI approaches can translate motor imagery and prosthetic limb movements into control of virtual avatars, eliminating the need for external sensors on the body \cite{wang2012self}. With the advancement of technology, we can expect future BCI-enabled Metaverse systems to feature lightweight, highly mobile BCI devices for interaction with Metaverse applications.

One of the central challenges in deploying HDT in the Metaverse is ensuring high accuracy in the data measured by integrated VR-BCI headsets, compared with data collected from conventional sensors. To address this challenge, future research may need to investigate the correlations and connections between brain signals and other human biological signals, such as the electrocardiogram (ECG) and electromyogram (EMG) \cite{lin2016delay}. Another challenge is real-time synchronization and communication between HDT avatars and the Metaverse users, which is essential for maintaining high-quality data within the Metaverse. These challenges are discussed further in Section \ref{subsec:real-time-commun}, respectively.

Once the correlations and relationships between signals among different brain lobes and the human's biological signals, e.g., ECG and EMG, are recognized, a wide range of applications for the Metaverse can emerge. Multimodal machine learning techniques, which recently advanced in processing large amounts of data from various sources or distributions \cite{zhang2020emotion, wen2020current}, can be a central component in a range of the Metaverse applications.
In \cite{karacsony2019brain}, the authors proposed an MI-BCI control framework that can work with multimodal signals, i.e., EEG and fNIRS. In particular, the proposed multimodal classifier based on a Convolutional Neural Network (CNN) can extract spatial and temporal features of both EEG signals and fNIRS images, thus resulting in higher classification accuracy. In \cite{bekele2016multimodal}, the authors proposed an interactive social platform that integrates eye gaze, EEG signals and peripheral psychophysiological signals of children with Autism Spectrum Disorders (ASD) in a VR setting. The aim of the study in \cite{bekele2016multimodal} is to understand the underlying factors that affect ASD children through emotion recognition tasks in a VR environment. As a result, potential future works can improve the emotion recognition abilities and eventual social functioning of children with ASD.

Although the works mentioned earlier achieved adequate performance with multimodal data in VR environments, incorporating such approaches into the Metaverse is still a significant research gap. The main difference between conventional VR settings and the Metaverse is that multiple HDT avatars are involved in the Metaverse, yielding more complicated interactive social systems between the HDT avatars. In other words, multiple `brains' of different individuals would be synchronized in the Metaverse. To fully understand the social connections and interactions between such HDT avatars, conventional approaches considering a single deep learning/machine learning classifier for an individual may fail to apply in a social HDT avatars setting \cite{zhang2021survey}. To address these challenges, transfer learning \cite{hajinoroozi2017deep} and meta-learning \cite{li2021model} can be applied. The transfer learning and meta-learning techniques share the same interest in utilizing underlying transferable features of the input signals, e.g., EEG, among different individuals. Once the learning models are trained, they can be transferred or directly applied to different HDT avatars with minimal fine-tuning processes. As a result, the Metaverse applications can only maintain a small number of learning models while guaranteeing high prediction accuracy of different tasks, e.g., emotion recognition and seizure prediction, compared with conventional machine learning approaches. This can reduce the maintenance, deployment, and scalability costs for the Metaverse applications.

\subsubsection{Wearable Sensors Technologies}
Wearable and portable technology has become ubiquitous in the current digital era. BCI technology can unlock a wide array of human sensing capabilities for a life-like HDT and can be further enhanced through additional wearable sensors. There are many platforms for wearable sensors such as smartphones, smartwatches, and smart clothing/jewellery \cite{wang2019survey}. These platforms utilise multiple types of sensors, such as pulse oximetry, electrodermal activity, global positioning system (GPS), inertial measurement unit (IMU), microphones, cameras, and temperature sensors (thermometers or thermistors). AI personalization and IoT data sharing can enhance the interpretation of these sensor measurements. These multimodal measurements and advanced technology culminate into a sensory information mapping system for emotions, cognitions, thoughts, and body movements. 

\textbf{Smartphones} are carried as an essential technology for everyday life. Over 80$\%$ of the world's population is estimated to own a smartphone \cite{berenguer2016smartphones}. Smartphones offer a portable platform for high-performing processing, network connectivity, and various onboard (or attached) sensors \cite{case2015accuracy}. Additionally, smartphones can be the gateway device for users to enter and interact with the Metaverse \cite{petrigna2022metaverse}. For the HDT-based Metaverse interaction, the smartphone can measure physical (steps and calendar/schedule logging), locational (GPS location and IoT geographical data), and social activity (AI personalisation and IoT social data) \cite{su2014activity}. This information can be used to formulate the activities and location of the HDT as a representation of the user in the Metaverse. 

\textbf{Smartwatches} is an emerging technology that is growing in popularity in daily life. A smartwatch is a wrist-worn computing device that is capable of communicating with other smartphones and computer devices \cite{cecchinato2015smartwatches}. Smartwatches are often used as a monitor fitness, an extension of a smartphone (phone calls, messaging, payment, and music), and as an assistive technology. Typically, smartwatches (such as Fitbit, Garmin, Apple, or Samsung watches) will carry a range of integrated sensors such as pulse oximetry, ECG, EMG, temperature, and IMU \cite{motahari2022health}. These sensors can actively measure physical exercise, basic emotional state (meditation and stress), health metrics (cardiovascular), and longitudinal activity (sleep, steps, and location) \cite{king2018survey}. Within the Metaverse, the HDT can utilise metrics to accurately represent real-world users' current bodily/mental state and physical activity. 

\begin{figure*}[t]
    \centering
    \includegraphics[width=0.9\linewidth]{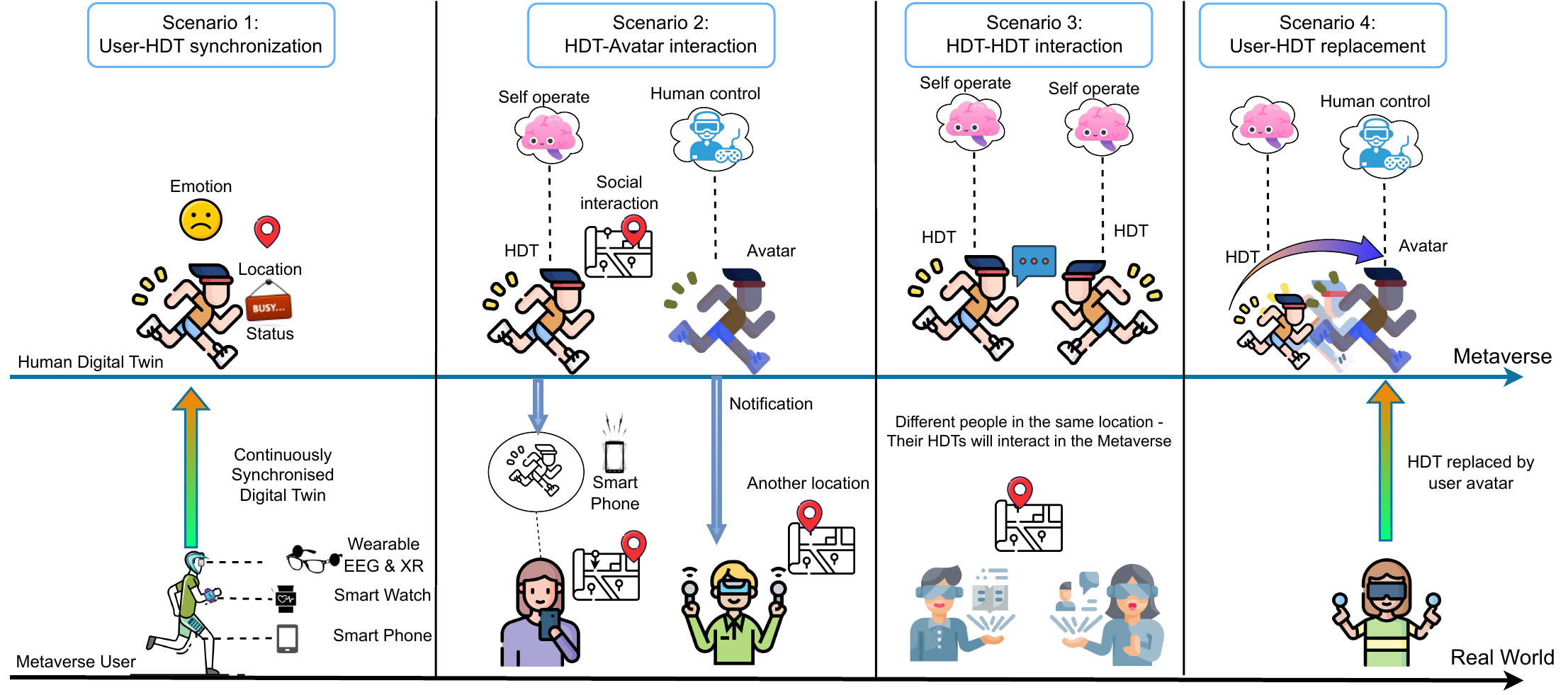}
    \caption{An illustration of the different scenarios where the Human Digital Twin (HDT) interacts with the Metaverse user, other avatars of Metaverse users, and other HDTs. Each scenario depicts the types of interactions that the HDT can engage in and the supporting technology to enable the interaction.}
    \label{fig:hdt-model}
\end{figure*}

\subsection{Human Digital Twin Interaction within the Metaverse}
\label{subsec:ApplicationHDT}
Fig.~\ref{fig:hdt-model} represents the different interaction scenarios we envision the HDT will engage within the Metaverse. The lifecycle of the HDT begins with the user leaving the Metaverse (Scenario 1: User-HDT synchronization). The HDT will replace the user's avatar and become of representation of the user when the user is present in the real-world. The HDT will continuously synchronize with the user state through various wearable technologies outlined in Fig.~\ref{fig:hdt-creation}. Within the Metaverse, the HDT can interact with the avatars of other Metaverse (Scenario 2: HDT-Avatar Interaction). The HDT will use technology, such as natural language models, to simulate a life-like interaction between an HDT and a Metaverse user. Alternatively, if two real-world users interact, their HDT will also interact (Scenario 3: HDT-HDT Interaction). This interaction will centre around the transference/sharing of information and ensuring that real-world users can inform each other with up-to-date information, similar to using social media. HDT's lifecycle ends when the real-world user enters the Metaverse (Scenario 4: User-HDT replacement). The Metaverse user's avatar would replace the HDT, and all of the HDT's activities would be synchronized with the Metaverse user. 

\subsubsection{User-HDT Synchronization}
The HDT addresses the key problem of discontinuity of a Metaverse user once they exit the Metaverse. In this scenario, when a user exits the Metaverse, the HDT will be activated to replace the user. As outlined in Section \ref{sec:HDTBCI}, the HDT can leverage several technologies, such as BCI, smartphones, smartwatches, and AI, to create a life-like extension and representation of the user within the Metaverse \cite{shengli2021human}. Through AI personalization and behaviour recognition, the HDT can interact with other avatars of users and HDTs within the Metaverse. The HDT can represent the current status of the real-world user through health and biometric tracking technology \cite{barricelli2019survey,okegbile2022human,martinez2019cardio}. The HDT can accurately reflect the real-world user's current status, activities, and interactions within the Metaverse.

\subsubsection{HDT-Avatar Interaction}
Social life-likeness is important when an HDT interacts with other Metaverse users' controlled avatars. While the HDT will likely have a unique status/identity within the Metaverse, the HDT must be life-like to supplement the user base of the Metaverse (similar to non-playable characters in video games) \cite{lim2012creating}. An HDT-to-avatar interaction can provide a naturalistic interface for Metaverse users to be informed on the status, activities, and locations of other users in the real-world \cite{duan2021metaverse}. By utilising natural language models, such as ChatGPT, the HDT can engage in authentic social conversations with Metaverse users \cite{taecharungroj2023can,rydell2022predictive}. This technology can further personalized natural language models by analysing imagined speech \cite{lee2022toward} and prior social engagements (social media or messaging) \cite{hodorog2022machine}. Overall, the HDT will act as a representation of the real-world user that is capable of engaging in meaningful social interactions with other Metaverse users.

\subsubsection{HDT-HDT Interaction}
HDT-to-HDT interaction is another distinctive form of interaction within the Metaverse. This type of interaction occurs when two real-world users interact within the real-world at the same location/area. In this situation, the HDTs will communicate in an esoteric manner to share information and update the real-world user on various life events, similar to social media information sharing \cite{osatuyi2013information}. This type of interaction shares similarities to the spatial location functions of the Snapchat social platform \cite{juhasz2018analyzing}. Within the Snapchat app, the spatial location of the users is visualised on a map with Bitmoji avatar representation. This spatial map is used to relay information, news, and other social events to groups of users occupying the same spatial area \cite{juhasz2018analyzing,wilken2021placemaking}. BCI and other wearable technologies can further enrich the behavior of HDT and shared information by measuring the thoughts and well-being of the users. With the growing usage of social media, HDT-to-HDT interaction can be a useful tool for the expedient social transference of information between two people.

\subsubsection{User-HDT Replacement}
The end of the HDT's lifecycle is when the real-world user re-enters the Metaverse (through smartphone, VR/AR/XR, or other devices). In this scenario, the user's avatar will replace the HDT and the Metaverse user can return to interacting within the Metaverse. During this scenario, the HDT can provide highlights and narratives akin to social media stories \cite{parry2022social}. Creating a narrative can enhance the acceptance of HDT as they are essential to informing and continuing the user's interaction/connection to the Metaverse \cite{lim2012creating}. Furthermore, the process of updating the user of the HDT's activities provides an incentive for other Metaverse users to interact with the HDT. The importance of the avatar's replacement of the HDT (over the HDT co-existing with the user) is to maintain the HDT's identity as an extension of the user. If the user co-exists with its own HDT in the Metaverse, it may create a sense of disembodiment or detachment between the HDT and the user \cite{mottelson2023systematic}. Therefore, the process of a user-controlled avatar replacing the HDT is paramount in facilitating a sense of embodiment and continuity for the user.

\subsection{Potential Challenges for the Development of the Human Digital Twin in the Metaverse}
\label{subsec:real-time-commun}
Another critical challenge in enabling the Metaverse applications with BCI is real-time synchronization and communication between the Metaverse users and their HDTs.
In the following, we focus on the communication aspect based on the two main perspectives that are (i) communications between BCI headsets and other infrastructures in the physical world and (ii) communications between human avatars and other avatars or technologies/virtual services in the Metaverse.
The first type of real-time communication is in the physical world, while the latter is in the Metaverse. Real-time communications in the physical world aim to provide robust and reliable connectivity for users with equipped VR/BCI headsets. The transmission of the brain signals over the network systems should meet low latency and error requirements. On the other hand, real-time communications in the Metaverse mainly occur between the users or their avatars with the environment, objects, or other avatars within the Metaverse. As a result, the requirement of real-time communications in the Metaverse is to achieve the continuity of user experience in the Metaverse where there is a parallel presence between the Metaverse and the real-world.

\subsubsection{Real-time Communications in the Physical World}
To achieve reliable and robust real-time communication between BCI headsets and other devices, the infrastructures that support wired/wireless communications play an essential role.
In conventional BCI systems, e.g., BCI2000 \cite{schalk2004bci2000}, real-time communications usually refer to scenarios where user brain signals are acquired with wired connections. Thanks to hardware development, wired connections are getting replaced by wireless connections with increasing mobility and reliability. Early research works in wireless BCI utilize Bluetooth for short-range communication between BCI headset and the processing unit, i.e., a computer \cite{lin2008development, he2015wireless, jafri2019wireless, lin2010real}. Although Bluetooth shows its capability in real-time communication, the communication range of Bluetooth is relatively short, i.e., from a few meters up to a range of ten meters. Further efforts to increase the communication range of the wireless BCI systems are reported in \cite{lin2008development, alshbatat2014eeg, rosenthal201925, navarro2004wearable, hieu2023enabling}. With significant increases in communication range \cite{lin2008development, alshbatat2014eeg, rosenthal201925} and joint computing-resource allocation \cite{hieu2023enabling}, wireless BCI shows its potential for enabling real-time communications between BCI headsets and other infrastructures in real-time at large scale.

As mentioned earlier, several works successfully investigated wireless BCI systems' connectivity and reliability. However, when large-scale systems include heterogeneous wireless devices and medium access schemes, the radio resources should be efficient management \cite{moioli2021neurosciences}. Specifically, Bluetooth and fiber connections may not always be available for BCI users because of coverage problems of such technologies. Such problems require new wireless access methods, radio resource allocation schemes, and broader bandwidth. In the following, we discuss the potential solutions for the problems mentioned above in wireless BCI systems.

To handle multiple requests and transmission of not only brain signals but also VR content over the wireless environment, Time Division Multiple Access (TDMA) can be an efficient solution. With TDMA, the time horizon is divided into multiple time slots, and the BCI users can communicate with the service providers or to each other in reserved time slots \cite{perfecto2020taming}. However, using time domain division also brings scheduling and data packet collision problems, thus making the TDMA-based systems hard to scalable. Recent advances in antenna design can enable many BCI users to use Multiple Input Single Output (MISO) and Multiple Input Multiple Output (MIMO) communications. For example, in MISO systems, the service provider can be a multi-antenna transmitter that can serve multiple users or multiple groups of users via Spatial Division Multiple Access (SDMA) \cite{wei2020prediction}. Advanced multiple-access methods can utilize the power domain to transmit VR applications and brain signals. For example, Non-orthogonal Multiple Access (NOMA) \cite{xiang2020noma} and Rate-splitting Multiple Access (RSMA) \cite{hieu2022virtual, mao2022rate} can be used to enhance data transmission rate, thus increasing the quality of service for the users. Besides advanced multiple access methods, 6G systems can utilize broadband communication techniques such as millimeter wave (mmWave) and Terahertz to enhance data transmission rate further. In such 6G systems, the data transmission rate is envisioned to be ten times faster than that of the 5G systems, making seamless experiences for data-demanded applications such as BCI-enabled Metaverse.

The techniques above and methods in wireless communications are promising for BCI-enabled Metaverse applications. However, the underlying theory behind such techniques and methods is based on Shannon's theory of wireless channel capacity. In other words, the data transmission rate of such methods cannot exceed the Shannon bound. Recent advances in machine learning and wireless communication techniques enable the transmission beyond Shannon bound with semantic compression and semantic communication \cite{qin2021semantic}. Unlike conventional data compression techniques, such as Shannon-Fano and Huffman coding, semantic compression is designed especially for machine-based communications in which intelligent machines only need specific semantic information to encode/decode the data successfully. On the other hand, semantic communication refers to using language and other symbolic systems to convey meaning between individuals or groups. Semantic communication involves transmitting words or signals and interpreting those words or signals within specific contexts for their intended meaning. Considering the brain signals as information that needs to be transmitted, we need further investigations on the semantic meaning of the brain signals, e.g., semantic reasoning of EEG signals \cite{klimesch1994episodic}, to design effective semantic communication frameworks for BCI-enabled Metaverse applications.

\subsubsection{Real-time Communications for the Human Digital Twin in The Metaverse}
Unlike real-time communications in the BCI/VR systems, real-time communications in the Metaverse refer to the scenario where users can interact with the Metaverse environment and their digital avatars in real-time. For this, the Metaverse applications should be able to provide highly user-driven embedded facilities such as real-time recommendations and individual support for the users through the virtual environment and digital avatars. For example, digital avatars can give valuable suggestions to users based on the analyzed brain signals. On the other hand, the quality of immersion of users in the Metaverse through VR/XR should also be highly considered. Our discussion about imagined speech communications \cite{lee2020neural}, adaptive VR/XR environment rendering \cite{cho2014bci}, error-related behaviors detection \cite{arico2017passive}, and HDT in the previous section can also be applied in this context. The embodiment of the human digital avatars, i.e., HDT, can be further enhanced by using a digital twin. With a digital twin, the human digital avatars can actively mirror their real-world counterparts through sensory data \cite{kritzler2017virtual}. In addition, digital twin avatars can simulate or predict potential user anomalous behaviors with reinforcement learning and deep learning in real-time \cite{moya2022digital, khan2022digital}. This functionality can also support other Metaverse-related interactions such as user collaboration, conferencing, presentations, and educational training/demonstrations \cite{dunleavy2014augmented}.

\section{Challenges, Emerging Applications, and Future Research Directions}
\label{sec:open-issues}
Despite the success in clinical trials and healthcare applications, there are still debates on using BCI technologies for commercial products. We discuss the open issues regarding the usability of BCI for the Metaverse, ethics, and security. We also review potential research directions of BCI toward a human-centric design for the Metaverse.

\subsection{Current Challenges}
\subsubsection{Hardware Development}
The first challenge comes from brain signal extraction for BCI applications. Although the portability and mobility of non-invasive BCI technologies enable commercial products, the brain signals extracted with such non-invasive BCI technologies through either dry or wet sensors usually come with a low signal-to-noise ratio (SNR) compared with the invasive methods. The reason is that the external sensors in non-invasive methods are further away from neuronal sources, plus noise, muscle contraction artifacts, and other tissue-related interference, making signals extracted less effective. On the contrary, invasive methods with the implanted electrode grids under the skull can provide less noisy and more reliable readings. However, the usability of invasive methods still faces critics and ethics concerns. 

The hardware and software capabilities remain open challenges toward the two distinct directions of the BCI methods. The invasive BCI research and development may focus on designing microchips and grids that can be effectively implanted under the skull. The recently funded companies such as NeuralLink \footnote{\url{https://neuralink.com/}} and BrainGate \footnote{\url{https://www.braingate.org/}} are operating toward this vision. However, such companies still develop their products based on clinical trials for patients with paralysis or animals. Apart from this direction, other companies such as Neurable \footnote{\url{https://neurable.com/about}} and OpenBCI \footnote{\url{https://openbci.com/about}} focusing on non-invasive BCI methods aim to develop portable, highly mobile BCI devices for daily use. Moreover, OpenBCI's Galea headset is a VR-BCI device that allows users to play games and interact with virtual environments through thoughts. We can expect potential Metaverse applications based on VR-BCI technology in the near future. However, the software development must be further developed due to the noisy nature of non-invasive BCI signals. The wearable weight and comfort of such devices should also be considered in design and development.

\subsubsection{Software Development}
The software development for BCI may heavily focus on extracting and monitoring brain signals. Understanding an individual's brain signals is yet a challenging task, let alone the neurodiversity among the population, age, race, and health. For example, the EEG signals from an individual may almost differ from the others in terms of amplitudes of the signals and the delayed responses to external events. As a result, one software or algorithm does not always fit all. Dealing with the neurodiversity between populations is still an open research issue \cite{zhang2021survey}. Few research works attempted to address this problem, but the existing methods are still limited to a few research participants \cite{zhang2017multi, eugster2014predicting, ji2016eeg, hieu2023enabling}. The common approaches of the above studies are additional feature extractions techniques \cite{zhang2017multi}, feature representation \cite{eugster2014predicting}, multimodal machine learning \cite{ji2016eeg}, and meta-learning \cite{hieu2023enabling}. The main goal of these works is to ensure that the trained machine-learning models can be applied to a new BCI subject without further user-specific training or calibration. Thus, this can significantly enhance the scalability and interoperability of the system.
For the large-scale BCI-enabled Metaverse systems, besides developing accurate signal processing schemes, resolving the neurodiversity of brain signals should be further investigated.

\subsubsection{Security and Privacy}
Recent studies showed that analyzing resting-state fMRI data \cite{ju2017early} and local field potentials \cite{mohammed2017toward} of the users with BCI can early reveal diseases such as Parkinson's and Alzheimer, respectively. In BCI-enabled Metaverse applications, the BCI headsets connected to the Internet also expose the users to potential privacy issues such as hackers, corporations, or government agencies that can track or even manipulate an individual's mental experience \cite{yuste2017four}.
Specifically, the advertisement-based applications in the Metaverse can gather BCI data from users to tailor their ads that target suitable individuals.
Similar to privacy issues of existing social media platforms such as Facebook and Twitter, in which the user data is the product of the social media platforms, the users' information can be sold to advertisement companies. To address this challenge, a decentralized Metaverse with a transparent consensus mechanism, e.g., blockchain \cite{zheng2017overview}, can prevent the data manipulation problems of a firm/company in a centralized Metaverse.

\subsubsection{Ethics}
Thanks to the rapid growth of machine learning and deep learning algorithms, much research has shown that using machine learning and deep learning enables highly accurate predictions and classifications on different BCI settings \cite{zhang2021survey}. Although such algorithms successfully achieved high performance, they are usually tricky or impossible to comprehend \cite{szegedy2013intriguing, arrieta2020explainable}. As a result, this introduces an unknown and unaccountable process between the neural pathways within the brain and the external technologies within the Metaverse \cite{drew2019ethics}. For example, the deep learning-aided auto-correction mechanism in imagined speech communication \cite{lee2022toward} may send unintended messages that the user is possessive about. To void this issue, implementing a BCI-enabled Metaverse system needs to prioritize ``how" and ``when" to send and/or collect imagined speech of the users. Toward an ethical solution for this, a collaborative project named BrainCom \cite{braincom}, funded by the European Union, is developing speech synthesizers with BCI technologies. Such technologies aim to vocalize the users' thoughts with ethical concerns accurately.

\subsection{Emerging Applications and Future Research Directions}
Although BCI has a long development history, integrating BCI into the Metaverse is still in its infancy. Thus, we expect that the interest of the industry and academy on this topic will expand in the following years. In this section, we discuss the emerging applications of BCI toward the Metaverse. Furthermore, we present several potential research directions.

\subsubsection{Integrated VR-BCI Devices}
Recent development of hardware and software for BCI applications in VR and XR enables the reduction in sizes and costs of the integrated VR-BCI devices. 
There are start-ups that are developing headsets allowing users to control the VR environment by using their EEG signals.
For example, Galea\footnote{\url{https://galea.co/}} and Cognixion ONE\footnote{\url{https://one.cognixion.com/}} are the headsets developed by OpenBCI and Cognixion, respectively, use from 6 to 8 dry EEG electrodes, in combination with transmission module, e.g., Bluetooth and WiFi, and eye tracking module. Galea is a VR-based headset, while Cognition ONE is an AR-based one, so these devices function very differently. Galea focuses on translating EEG signals into digital commands for VR games and applications. On the other hand, Cognixion ONE uses a combination of EEG signals, eye movement, and facial expressions to control digital devices and interact with augmented environments. 
Overall, both Galea and Cognixion ONE are innovative and exciting products that are pushing the boundaries of brain-computer interface technology. However, they have different strengths and applications, so their choice will depend on the user's needs and preferences.
For example, Galea is more promising for gaming applications in the BCI-enabled Metaverse, where the users can control their players by thoughts, and Cognition ONE is more appropriate for healthcare Metaverse applications. 

\subsubsection{Multitasking in the Metaverse}
Most of the current machine learning and deep learning approaches for BCI applications are task-specific, meaning that a machine learning model is associated with an individual, given a specific demand, e.g., emotional detection and seizure prediction \cite{zhang2021survey}. This approach is suitable for the classifiers to be deployed at the user site, e.g., a pre-installed software in the headset. However, when it comes to multitasking applications, e.g., combined imagined speech and emotional recognition, the task-specific classifier/software may fail or downgrade the performance. 
For example, in some Metaverse applications such as virtual fighting, fitness, and dance gaming, the multisensory data from EEG, facial expression, and emotional states can be jointly exploited to enable secure imagined conversations between groups of users and reduce potential motion sickness induced by fast-moving scenes.
To fully exploit the multisensory data in such scenarios, multimodal machine learning approaches, which we described in Section \ref{sec:HDTBCI}, can be a suitable approach. With multimodal machine learning, we can expect only one machine learning model to assist the user in various tasks, from emotional recognition to imagined speech communication, without requiring changes or upgrades in software and hardware.

\subsubsection{Potential Research Directions}
Regarding the technical challenges and usability of the BCI-enabled Metaverse concept, different aspects of this new concept must be further investigated and studied. The following discusses the potential research directions for the BCI-enabled Metaverse systems.
\textbf{Machine Learning for processing heterogeneous datasets:}
In recent years, the notable trend in BCI is applying advanced machine learning techniques for analyzing brain signals. Similarly, machine learning techniques will be a significant component in the BCI-enabled Metaverse system. Unlike conventional problems in BCI research, which extensively focus on designing highly accurate classifiers for specific tasks \cite{zhang2021survey}, the integration of BCI into Metaverse brings new challenges. 
One of the most notable is addressing the complexity of the system processing multiple sources of human brain signals. As shown in early findings \cite{zhang2017multi, ji2016eeg, hieu2023enabling}, the participation of multiple users in a task yields a degraded performance for the classifier due to the neurodiversity among the users. The neurodiversity suggests that the brain signals such as EEG or fMRI are highly individual and different among users based on gender, age, and other factors \cite{meng2017predicting}. As a result, a classifier trained with one dataset, e.g., EEG, of a person does not work well with one another. 

Conventional approaches with machine learning require different classifiers for different people, thus resulting in inefficient computing and poor usability. 
To address this problem, further analysis on the brain signals \cite{zhang2017multi, ji2016eeg} or advanced machine learning algorithm, e.g., meta-learning \cite{hieu2023enabling}, can be applied. As a result, only one classifier can be used across the users without degrading the performance. Although a few early works address the neurodiversity problem, the large number of data generated from human activities, including human brain signals, poses new challenges. The virtual environments in the Metaverse also raise technical concerns about combining virtual environments' constraints, e.g., motion sickness and delay, into creating effective machine learning models. The multiple modalities of the data sources make it hard to understand and analyze the informative features. The multimodal machine learning techniques we discussed in the previous sections can be a potential solution. Apart from multimodal approaches, the success of attention-based machine learning models, e.g., Transformer \cite{vaswani2017attention}, in understanding complex problems, ranging from visual scenes to language understanding, make it a potential candidate for tacking the multiple modalities of the evolved data in the Metaverse. The attention-based mechanisms with designed attention vectors make the machine learning models can pay attention to the most valuable parts of the data, thus making the data extraction process more effective. In addition, the applications across the virtual worlds in the Metaverse can share similar features, e.g., user behaviors in applications such as e-commerce. Therefore, we can better exploit the valuable features and optimize the machine learning models by learning transferable features of the virtual worlds. In such scenarios, transfer learning techniques are commonly used \cite{hajinoroozi2017deep}.

\textbf{Human Digital Twin for Maintaining Continuity in the Metaverse:}
As discussed in Section ~\ref{sec:human-digital-twin}, HDT will play a key role in helping us better understand the virtual embodiment of the human body in the Metaverse. In future Metaverse applications, HDT should provide intelligent interfaces for digital avatars by using the brain signals and leveraging 3D visual effects of the human body, e.g., facial expression and body pose, to construct individual avatars. 
As a result, the continuity of the Metaverse can be maintained in which the HDT can auto-generate or self-operate in the virtual environment with less control from humans. The human users can join the Metaverse and replace the HDT by their avatars when wearing VR/XR headsets (see Fig.~\ref{fig:hdt-model}).
To this end, the brain signals can be mixed with other human data such as body motion with physical engines \cite{winkler2022questsim, yi2022physical}, and facial expression \cite{raj2021pixel} to construct the most realistic and personalized HDT in the Metaverse.
This can be a new research area that lies at the intersection of neuroscience, e.g., using human biological data, and 3D avatar construction, e.g., using facial expression, limb movement, and kinematics study of the human body.

\section{Conclusion}
\label{sec:conclusion}
This survey provides an in-depth overview of non-invasive BCI technologies and their potential applications in the Metaverse. With BCI-enabled applications, the Metaverse is expected to be highly personalized and customized to individual needs. Furthermore, the users can interact with the virtual environments with a limited number of sensors, such as kinematics sensors and handheld devices. We also discussed a novel concept of HDT, where the twin representations of the users in the virtual worlds can be developed using a digital twin. Lastly, we discussed the open issues, including security, privacy, hardware and software capability. The potential research directions were also covered. Alternatively, the survey outlines the initial steps for the potential research area evolved at the intersection of BCI and the Metaverse technologies, e.g., VR/XR, 3D environment construction, and real-time communication and synchronization.

\bibliographystyle{IEEEtran}
\bibliography{refs}

\begin{IEEEbiography}[{\includegraphics[width=1in,height=1.25in,clip,keepaspectratio]{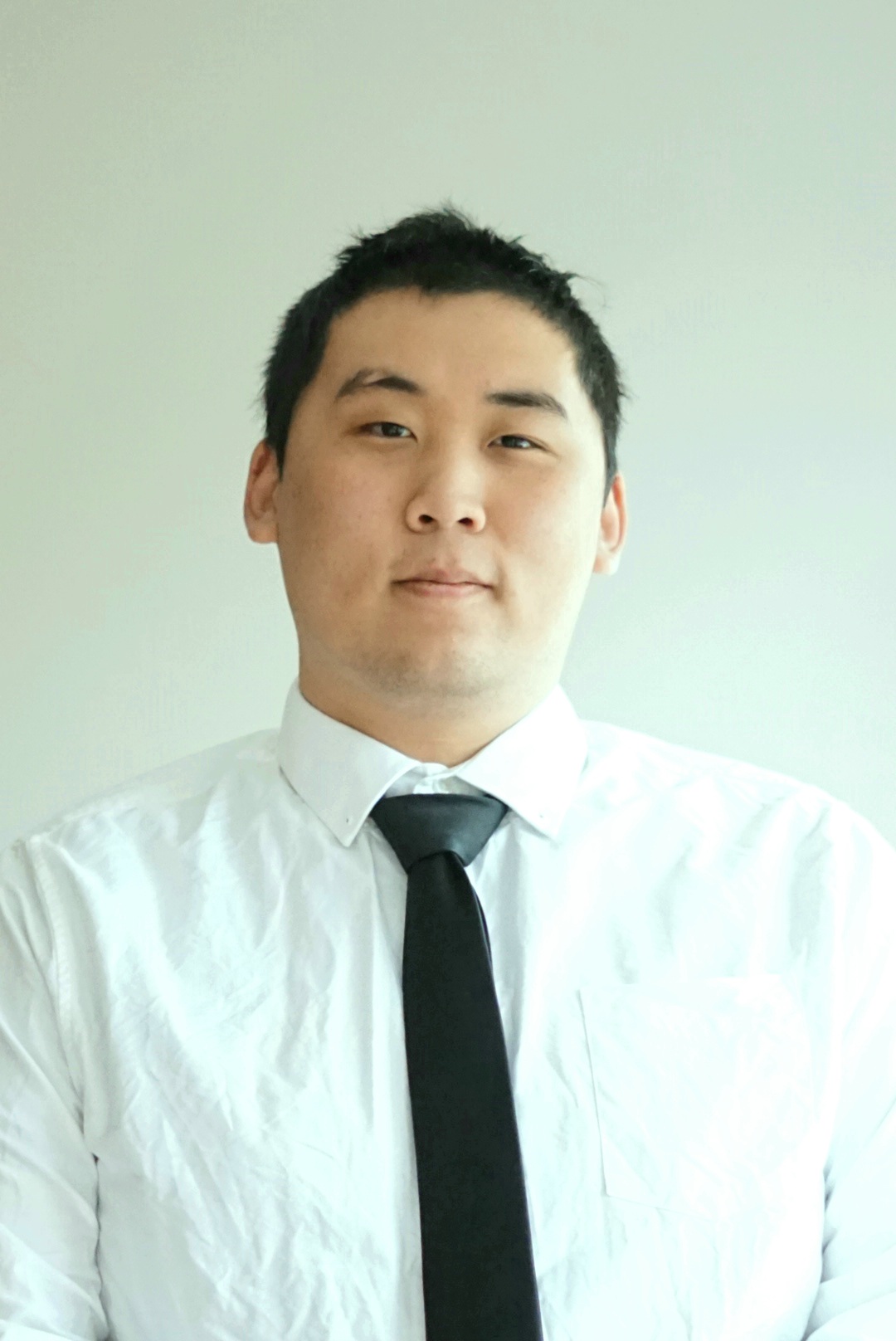}}]{Howe Yuan Zhu} received his Ph.D. degree in computer science at the University of Technology Sydney in 2022. He received his BSc in Biomedical Engineering at the University of Sydney in 2017. He is currently a research associate at the University of Technology Sydney and a core GrapheneX-UTS Human-centric Artificial Intelligence Centre (HAI) member. His current research interests are XR/VR/AR, brain-computer interfaces, robotics, artificial intelligence in robotics, and human-computer interaction. He is actively exploring translational applications of wearable XR technologies for medical applications and understanding human factors for assistive technologies.\end{IEEEbiography}

\begin{IEEEbiography}[{\includegraphics[width=1in,height=1.25in,clip,keepaspectratio]{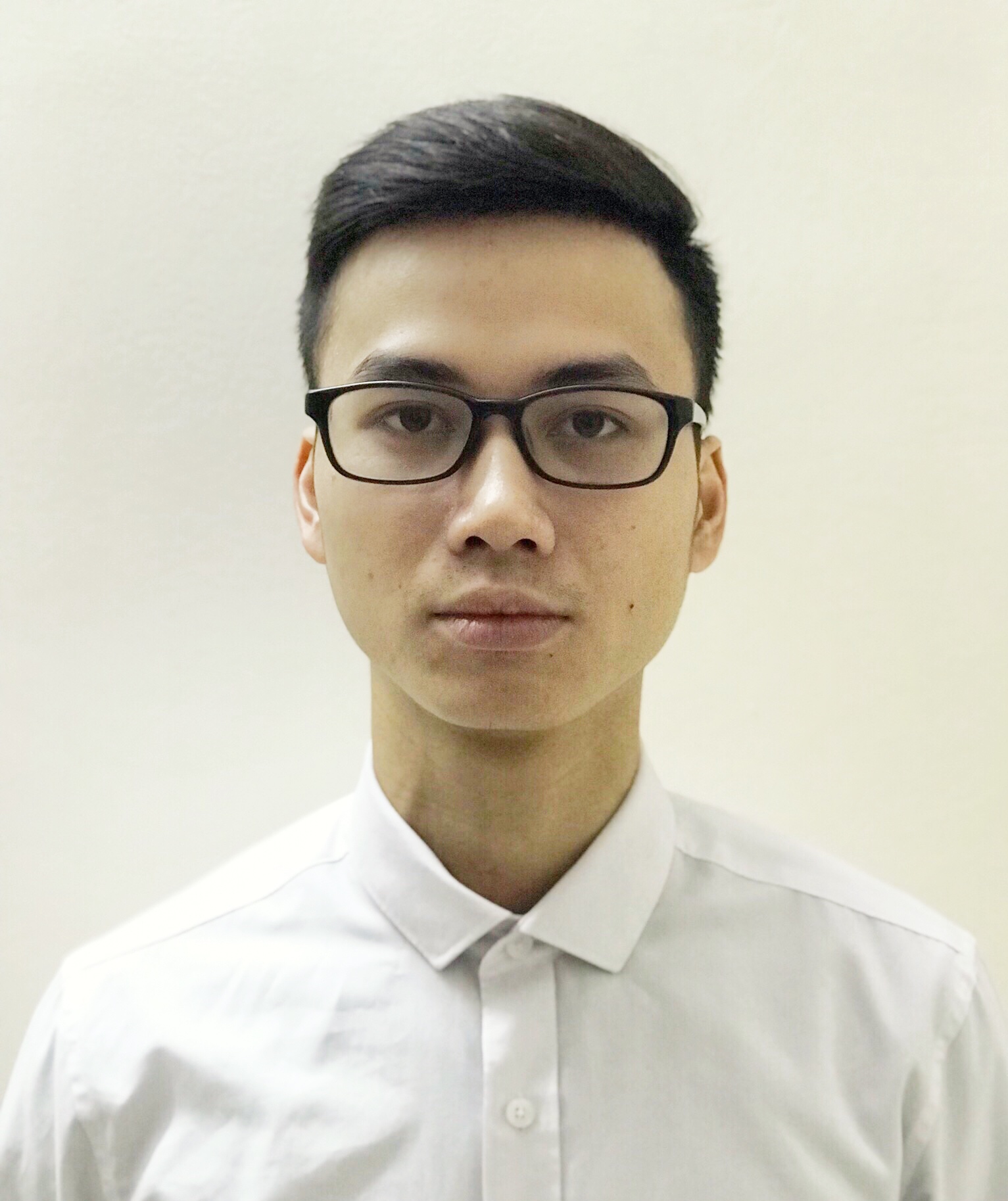}}]{Nguyen Quang Hieu} received the B.E. degree in Hanoi University of Science Technology, Vietnam in 2018. He is currently a Ph.D. student at School of Electrical and Data Engineering, University of Technology (UTS), Sydney, Australia. Before joining UTS, he was a research assistant at School of Computer Science and Engineering, Nanyang Technological University, Singapore. His research interests include wireless communications, network optimization, reinforcement learning, and deep learning..\end{IEEEbiography}

\begin{IEEEbiography}[{\includegraphics[width=1in,height=1.25in,clip,keepaspectratio]{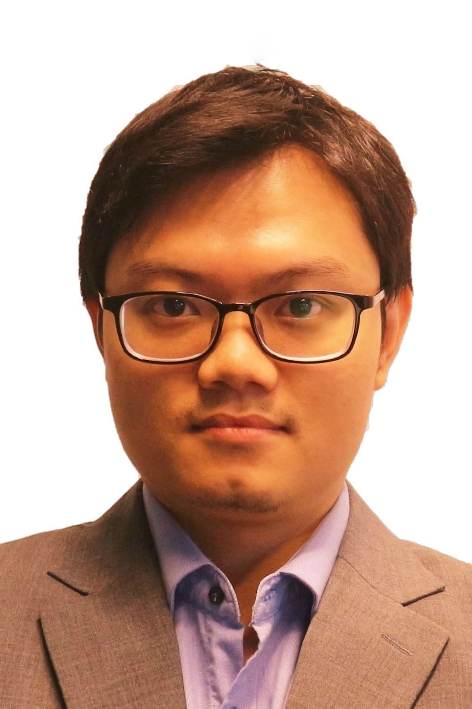}}]{Dinh Thai Hoang} (M’16, SM’22) is currently a faculty member at the School of Electrical and Data Engineering, University of Technology Sydney, Australia. He received his Ph.D. in Computer Science and Engineering from the Nanyang Technological University, Singapore 2016. His research interests include emerging wireless communications and networking topics, especially machine learning applications in networking, edge computing, and cybersecurity. He has received several precious awards, including the Australian Research Council Discovery Early Career Researcher Award, IEEE TCSC Award for Excellence in Scalable Computing for Contributions on “Intelligent Mobile Edge Computing Systems” (Early Career Researcher), IEEE Asia-Pacific Board (APB) Outstanding Paper Award 2022, and IEEE Communications Society Best Survey Paper Award 2023.  He is currently an Editor of IEEE TMC, IEEE TWC, IEEE TCCN, IEEE TVT, and IEEE COMST.\end{IEEEbiography}

\begin{IEEEbiography}[{\includegraphics[width=1in,height=1.25in,clip,keepaspectratio]{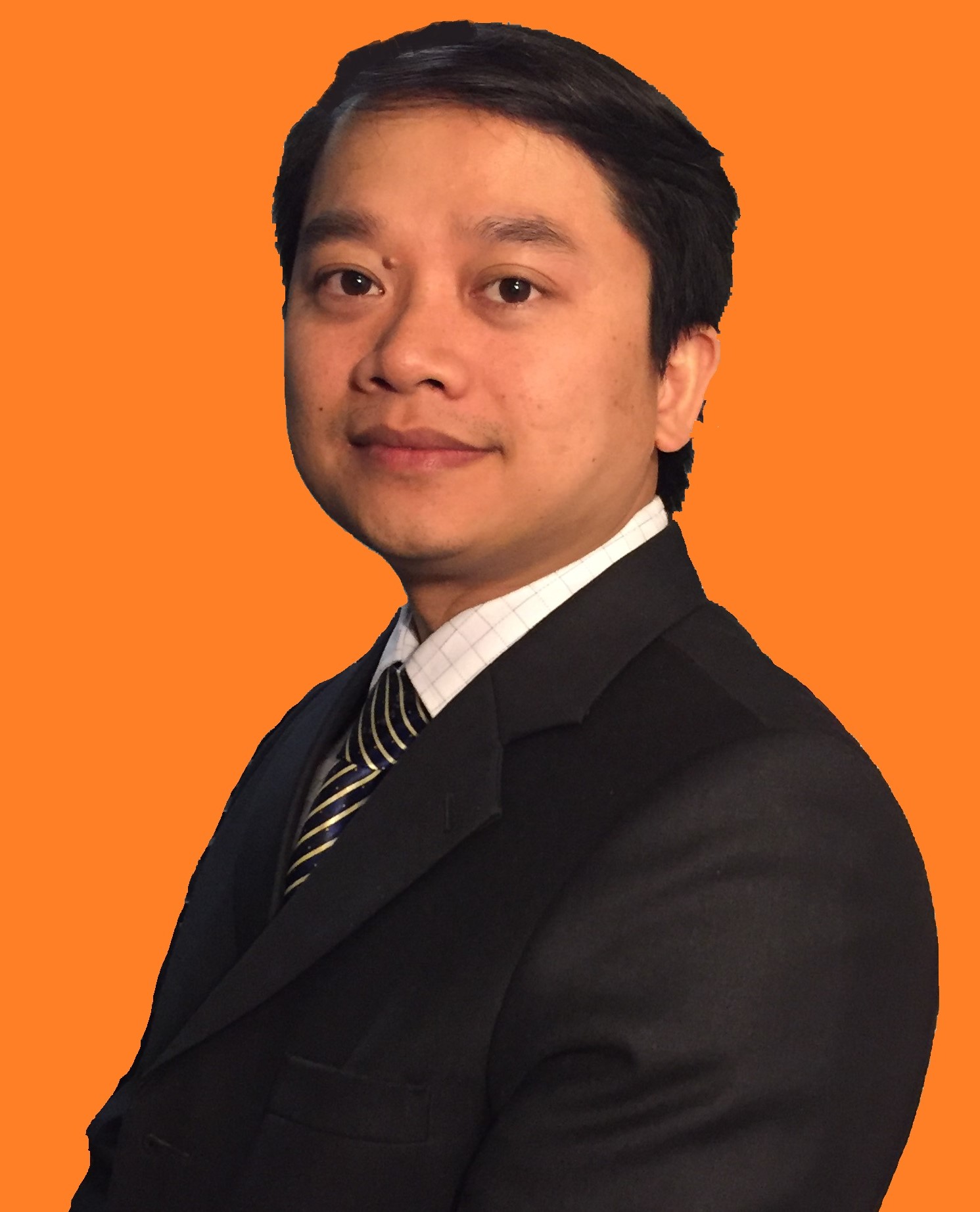}}]{Diep N. Nguyen}
 (Senior Member, IEEE) received the M.E. degree in electrical and computer engineering from the University of California at San Diego (UCSD), La Jolla, CA, USA, in 2008, and the Ph.D. degree in electrical and computer engineering from The University of Arizona (UA), Tucson, AZ, USA, in 2013. He is currently a faculty member with the Faculty of Engineering and Information Technology, University of Technology Sydney (UTS), Sydney, NSW, Australia. Before joining UTS, he was a DECRA Research Fellow with Macquarie University, Macquarie Park, NSW, Australia, and a Member of the Technical Staff with Broadcom Corporation, San Jose, CA, USA, and ARCON Corporation, Boston, MA, USA, and consulting the Federal Administration of Aviation, Washington, DC, USA, on turning detection of UAVs and aircraft, and the U.S. Air Force Research Laboratory, USA, on anti-jamming. His research interests include computer networking, wireless communications, and machine learning application, with emphasis on systems’ performance and security/privacy. Dr. Nguyen received several awards from LG Electronics, UCSD, UA, the U.S. National Science Foundation, and the Australian Research Council. He has served on the Editorial Boards of the IEEE Transactions on Mobile Computing, IEEE Communications Surveys \& Tutorials (COMST), IEEE Open Journal of the Communications Society, and Scientific Reports (Nature's).\end{IEEEbiography}

\begin{IEEEbiography}[{\includegraphics[width=1in,height=1.25in,clip,keepaspectratio]{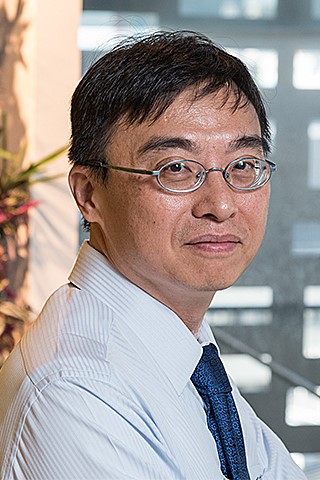}}]{Distinguished Professor Chin-Teng Lin} received a Bachelor’s of 
Science from National Chiao-Tung University (NCTU), Taiwan in 1986, and holds 
Master’s and PhD degrees in Electrical Engineering from Purdue University, USA
, received in 1989 and 1992, respectively. 

He is currently a distinguished professor and Co-Director of the Australian Artificial Intelligence Institute 
within the Faculty of Engineering and Information Technology at the University 
of Technology Sydney, Australia. He is also an Honorary Chair Professor of 
Electrical and Computer Engineering at NCTU. For his contributions to 
biologically inspired information systems, Prof Lin was awarded Fellowship 
with the IEEE in 2005, and with the International Fuzzy Systems Association (
IFSA) in 2012. He received the IEEE Fuzzy Systems Pioneer Award in 2017. He 
has held notable positions as editor-in-chief of IEEE Transactions on Fuzzy 
Systems from 2011 to 2016; seats on Board of Governors for the IEEE Circuits 
and Systems (CAS) Society (2005-2008), IEEE Systems, Man, Cybernetics (SMC) 
Society (2003-2005), IEEE Computational Intelligence Society (2008-2010); 
Chair of the IEEE Taipei Section (2009-2010); Chair of IEEE CIS Awards 
Committee (2022, 2023); Distinguished Lecturer with the IEEE CAS Society (2003-
2005) and the CIS Society (2015-2017); Chair of the IEEE CIS Distinguished 
Lecturer Program Committee (2018-2019); Deputy Editor-in-Chief of IEEE 
Transactions on Circuits and Systems-II (2006-2008); Program Chair of the IEEE 
International Conference on Systems, Man, and Cybernetics (2005); and General 
Chair of the 2011 IEEE International Conference on Fuzzy Systems.

Prof Lin is the co-author of Neural Fuzzy Systems (Prentice-Hall) and the 
author of Neural Fuzzy Control Systems with Structure and Parameter Learning 
(World Scientific). He has published more than 425 journal papers including 
about 200 IEEE journal papers in the areas of neural networks, fuzzy systems, 
brain-computer interface, multimedia information processing, cognitive 
neuro-engineering, and human-machine teaming, that have been cited more than 
34,000 times. Currently, his h-index is 90, and his i10-index is 401.\end{IEEEbiography}

\end{document}